\documentclass[apj,twocolumn,numberedappendix,twocolappendix,appendixfloats,tighten]{openjournal}
\usepackage{aas_macros}
\usepackage{amsmath}
\usepackage[usenames,dvipsnames]{xcolor}
\usepackage[caption=false]{subfig}
\usepackage{hyperref}
\hypersetup{colorlinks=true,linkcolor=blue,citecolor=blue,filecolor=blue,urlcolor=blue}
\usepackage{graphicx}

\allowdisplaybreaks[1]

\newcommand{\App}[1]{Appendix~\ref{#1}}

\newcommand{\Eq}[1]{Eq.~(\ref{#1})}

\newcommand{\Fig}[1]{Fig.~\ref{#1}}
\newcommand{\Sec}[1]{Section~\ref{#1}}

\def\arepo    {{\sc Arepo}}
\def\athena    {{\sc Athena++}}
\def\ramses    {{\sc Ramses}}

\def\JO        {{JO18}}
\def\runonemom {{\sc 1-mom}}
\def\runtwomom {{\sc 2-mom}}
\def\runnocr {{\sc NoCR}}

\def\ccr {\tilde c}           
\def\ccrmin {\tilde c_{\rm min}}  
\def\cms {\rm cm \, s^{-1}}           
\def\dtcr {\Delta t_{\rm c}}           
\def\dtmhd {\Delta t_{\rm MHD}}           
\def\Ecr {e_{\rm c}}           
           
\def\Ecrunits {{\rm erg \, cm^{-3}}}
           
\def\Fcr {F_{\rm c}}

\def\kcr {\kappa}           
\def\kcrunits {\rm cm^2 \, s^{-1}}           
\def\Nsc {N_{\rm sc}}           
\def\Pcr {P_{\rm c}}           
\def\Pgas {P_{\rm th}}

\def\va {u_{\rm A}}           
\def\vavec {\vec{u}_{\rm A}}           
\def\vecB {\vec{B}}
\def\vgas {{u}}           

\mathchardef\mhyphen="2D

\long\def\symbolfootnote[#1]#2{\begingroup%
\def\thefootnote{\fnsymbol{footnote}}\footnote[#1]{#2}\endgroup}

\begin{document}

\title[Two-moment CR transport in RAMSES]{Two-moment cosmic ray transport in RAMSES\vspace{-15mm}}

\author{
Joki Rosdahl$^{1*}$, Yohan Dubois$^2\dag$, Benoît Commerçon$^1$, \\ Nimatou Diallo$^2$, Nai Chieh Lin$^1$, and Alexandre Marcowith$^{3}$}
\thanks{$^*$E-mail: \href{mailto:karl-joakim.rosdahl@univ-lyon1.fr}{karl-joakim.rosdahl@univ-lyon1.fr}}
\thanks{$^\dag$E-mail: \href{mailto:dubois@iap.fr}{dubois@iap.fr}}

\affiliation{$^1$ Université de Lyon, Université Lyon1, ENS de Lyon, CNRS, Centre de Recherche Astrophysique de Lyon UMR5574, F-69230, Saint-Genis-Laval, France}
\affiliation{$^2$ Institut d’Astrophysique de Paris, Sorbonne Université, CNRS, UMR 7095, 98 bis bd Arago, 75014 Paris, France}
\affiliation{$^3$ Laboratoire Univers et Particules de Montpellier, CNRS/Université de Montpellier, Montpellier, France}



\begin{abstract}
Cosmic rays (CRs) are an important source of feedback  in a variety of astrophysical contexts. Magneto-hydrodynamical (MHD) simulations treating CRs as a fluid have shown that how their feedback operates is strongly dependent on their transport properties such as diffusion and streaming. In this paper we introduce the numerical implementation, in the adaptive-mesh-refinement MHD code \ramses, of the grey two-moment formulation of CR fluid dynamics, which follows the energy density and its associated three-dimensional flux. This method is tested for CR diffusion, streaming, and advection in a series of multi-dimensional tests including shocks to check the robustness and stability of this numerical two-moment CRMHD solver. We finally use the new two-moment CR implementation in a complex simulation of an isolated galactic disc producing  galaxy-wide outflows launched by small-scale supernova explosions, and compare it with a previously existing one-moment formulation in the same code.
\end{abstract}

\section{Introduction} \label{sec:intro}

Cosmic rays (CRs) are a population of supra-thermal particles produced in astrophysical shocks corresponding to various sources, such as supernova remnants (SNRs) and novae, magnetospheres of pulsars, superbubbles in massive star clusters and, at the highest energies, jets and shocks associated with active galactic nuclei and galaxy clusters. Among these sources, SNRs remain the most widely recognised candidates for supplying the bulk of galactic CRs, as their energetics and frequency are sufficient to explain the measured CR spectrum on Earth~\citep{Gabici19}. 
For sufficiently strong shocks and through the process of diffuse shock acceleration~\citep[e.g.][]{Bell78,Drury83,Marcowith16}, the CR energy extracted from the shock kinetic energy is of the order of 10\%~\citep[e.g.][]{Helder12,Caprioli14}.
Due to their close coupling with magnetic fields, CR protons exert an effective pressure on the thermal plasma, which can provide a source of feedback for the gas on galactic scales~\citep[see][for a recent review]{Ruszkowski23}. 


The transport of CRs through magnetized plasma is governed by their interactions with the turbulent magnetic field. Because CRs gyrate around field lines, their motion is initially guided along the ordered magnetic field direction. However, small-scale fluctuations in the field scatter their pitch angle, effectively randomizing the direction of propagation. These fluctuations may be externally driven by turbulence, or they may be generated self-consistently by CRs themselves. In particular, CRs drifting faster than the local Alfv\'en speed excite the so-called gyro-resonant streaming instability~\citep{Kulsrud69}, which amplifies magnetic perturbations that scatter the CRs. This self-regulation enforces a CR bulk drift speed of order the Alfv\'en velocity relative to the gas in addition to plasma (CR) heating (resp. loss), a process referred to as CR streaming.
The turbulent structure of galactic magnetic fields introduces an additional layer of complexity. Magnetic field lines are tangled by turbulence, and CRs tied to these lines undergo a form of random walk, resulting in transport that appears as diffusion on large scales. The effective transport is highly anisotropic: rapid along field lines, but much slower perpendicular to them. Describing this behavior, thus, requires modeling anisotropic diffusion coefficients that can vary with environment and magnetic geometry~\citep{Shalchi09}.
These processes not only control how CRs redistribute their pressure support across galactic environments, but also determine where they ultimately deposit their energy and momentum.

A consequence of this transport is that CRs are displaced away from their production sites in regions of high gas density, such as star-forming clouds and SNRs, into more diffuse environments. Since CR losses, driven by Coulomb and hadronic interactions with the background plasma, operate most efficiently in dense gas, this outward redistribution significantly reduces the overall energy loss rate of the CR population. As a result, CRs can persist for long periods and accumulate with a galactic energy density that reaches rough equipartition with the thermal, turbulent, and magnetic components, with energy densities typically around $\sim 1\,\rm eV\,cm^{-3}$~\citep{Boulares90}. This ability to act as a long-lived, mobile reservoir of non-thermal energy makes CRs a potentially important driver of feedback on galactic scales.


Interest in CRs as a source of astrophysical feedback has grown considerably, especially following hydrodynamical simulations that highlighted their potential to drive galactic winds and regulate star formation. Early studies employed simplified models of CR transport via isotropic diffusion or streaming~\citep{Jubelgas08, Uhlig12, Booth13, Salem14}. More recent work incorporating anisotropic diffusion and streaming has refined our understanding of CR-driven feedback, demonstrating that the magnitude of diffusion and streaming critically influences wind launching and galactic-scale regulation~\citep{Hanasz13,Girichidis16, Girichidis18, Simpson16, Pakmor16Winds, Ruszkowski17, Chan19, Dashyan20, Hopkins21, Hopkins21FeedbackTransport, Farcy22, Farcy25, MartinAlvarez23, Rodriguez24}.
Beyond galactic-scale feedback, CRs are believed to play a key role in several other processes: enhancing momentum deposition in SNRs~\citep{Diesing18, Rodriguez22}, shaping the density and turbulence structure of the interstellar medium~\citep{Commercon19, Dubois19, Nunez22, Simpson23, Sampson25}, and potentially contributing to the reheating of the interstellar medium \citep{Wiener13} or of galaxy cluster cores~\citep{Sijacki08, Ruszkowski17Cluster, Ehlert18, Yang19, Su21, Beckmann22}.

Accurately capturing these processes in numerical simulations requires robust models of CR transport. In the widely used ``one-moment'' formulation, the CR energy density is evolved via an advection–diffusion equation, with transport coefficients encoding diffusion and streaming. However, this approach faces significant numerical challenges. Diffusion is governed by a parabolic equation, which imposes a timestep constraint scaling as $\Delta t_\kappa = \Delta x^2 / (2\kappa)$, where $\Delta x$ is the volume element width and $\kappa$ is the diffusion coefficient. This timestep is often orders of magnitude smaller than the hydrodynamical timestep, $\Delta t_{\rm h} \approx \Delta x / v_{\rm max}$, set by the Courant condition. Streaming makes matters worse: because the streaming velocity is discontinuous at extrema of the CR energy density, the corresponding stability condition scales as $\Delta t_{\rm s} \propto \Delta x^3$. At high resolution, these restrictions render explicit schemes impractical. To circumvent this, several strategies have been proposed, including subcycling, implicit or semi-implicit solvers~\citep{Sharma11, Dubois16, Dubois19, Pakmor16}, or artificial smoothing of the streaming velocity~\citep{Sharma10}. In addition, ensuring correct entropy production requires specialized flux limiters~\citep{Sharma07}. Each of these strategies adds computational cost and algorithmic complexity.


A promising alternative is the two-moment formulation of CR transport, derived directly from the Vlasov equation. Instead of evolving only the CR energy density, this method evolves both the energy density and the energy flux. The system of equations is then hyperbolic, resembling the moment equations of radiative transfer. The timestep is limited by a Courant-like condition involving the (reduced) speed of light rather than the much smaller diffusive timestep~\citep{Snodin06, Jiang18, Thomas19, Chan19, Thomas21, Hopkins22}. In the limit where the CR flux relaxes to equilibrium rapidly compared to the dynamical timescale, the two-moment formulation recovers the classical one-moment equation. Thus, it provides a physically consistent extension of the one-moment method, mitigating the stiffness of a parabolic formulation and enabling significantly larger timesteps. Conceptually, it bridges the gap between kinetic and fluid descriptions, offering both computational efficiency and improved accuracy.


In this paper, we present an implementation of the two-moment CR transport equations in the \ramses{} code. We describe the theoretical derivation and numerical implementation of the method in Section~\ref{sec:methods}. In Section~\ref{sec:tests}, we validate it against a suite of multidimensional test problems, quantifying its accuracy, stability, and robustness. In Section~\ref{sec:galaxy}, we apply the scheme to the launching of a CR-driven galactic wind in an isolated disc configuration, similar to those used in previous studies~\citep{Dashyan20, Farcy22}, to demonstrate the applicability of the approach in a realistic context. Finally, in Section~\ref{sec:conclusions}, we summarize our results and discuss prospects for future applications in galaxy formation and feedback studies.

\section{Numerical method} \label{sec:methods}

The two-moment formulation of the CR equations and its numerical implementation in \ramses{}~\citep{Teyssier02} that we introduce here,  closely resembles that of~\citet[][\JO{} hereafter]{Jiang18}.  We refer the reader to~\cite{Thomas19} and \cite{Hopkins22} for a recent complete derivation of this two-moment formulation.

\subsection{The two-moment equations of CRMHD}

The full set of cosmic-ray magneto-hydrodynamical (CRMHD) equations is
\begin{align}
& \label{eq:mass} \frac{\partial \rho}{\partial t} + \vec{\nabla}  . (\rho \vec{u}) = 0 \,, \\
& \label{eq:momentum} \frac{\partial (\rho \vec{u})}{\partial t} + \vec{\nabla}  . \left(\rho \vec{u} \otimes \vec{u} + P - \frac{\vec{B} \otimes \vec{B}}{4\pi} \right) = -(\vec \nabla  P_{\rm c})^* \,, \\
& \label{eq:energy} \frac{\partial e}{\partial t} + \vec{\nabla}  . \left( (e + P) \vec{u} - \frac{(\vec{B}  . \vec{u}) \vec{B}}{4\pi} \right) = -\vec{u}  . \vec{\nabla} P_{\rm c} + \mathcal{L}_{\rm rad} \,, \\
& \label{eq:magnetic} \frac{\partial \vec{B}}{\partial t} - \vec{\nabla} \times (\vec{u} \times \vec{B}) = 0 \,, \\
& \label{eq:ecr} \frac{\partial e_{\rm c}}{\partial t} + \vec{\nabla}  . \vec{F}_{\rm c} = \vec{u}  . \vec{\nabla} P_{\rm c} + \mathcal{L}_{\rm rad,c} \nonumber \\
&\qquad - \vec{u}_{\rm s}  . 3(\gamma_{\rm c} - 1) \sigma_\kappa \left( \vec{F}_{\rm c} - (\vec{u} + \vec{u}_{\rm s})(e_{\rm c} + P_{\rm c}) \right) \,, \\
& \label{eq:fecr} \frac{1}{c^2} \frac{\partial \vec{F}_{\rm c}}{\partial t} + \vec{\nabla}.(\mathcal{D}_{\rm c} e_{\rm c}) = -\sigma_\kappa \left( \vec{F}_{\rm c} - (\vec{u} + \vec{u}_{\rm s})(e_{\rm c} + P_{\rm c}) \right) \,,
\end{align}
where $\rho$ is the gas mass density, $\vec{u}$ the gas velocity, $\vec{B}$ the magnetic field, and $e = e_{\rm th} + 0.5\rho u^2 + B^2/(8\pi)$ the total plasma energy density. The plasma pressure is $P = P_{\rm th} + B^2/(8\pi)$. The thermal and CR energy densities are denoted $e_{\rm th}$ and $e_{\rm c}$, with ideal equations of state: $e_{\rm th} = (\gamma - 1) P_{\rm th}$ and $e_{\rm c} = (\gamma_{\rm c} - 1) P_{\rm c}$, and adiabatic indices $\gamma = 5/3$ and $\gamma_{\rm c} = 4/3$, respectively.
Here, $\vec{F}_{\rm c}$ is the CR energy flux in the lab frame\footnote{The lab-frame flux $\vec{F}_{\rm c}$ with respect to the comoving fluid-frame flux $\vec{F}_{\rm c,\rm com}$ is defined as $\vec{F}_{\rm c}=\vec{F}_{\rm c,\rm com}+u(\Ecr+\Pcr)$}, and the streaming velocity is $\vec{u}_{\rm s} = -\vec{u}_{\rm A} \, {\rm sign}(\vec{B}  . \vec{\nabla} e_{\rm c})$, with $\vavec = \vec{B}/\sqrt{4\pi \rho}$ the Alfvén velocity. The equation on the CR flux contains an energy tensor $\mathcal{D}_{\rm c}$ that we take equal to the $\mathcal{D}_{\rm c}=\mathbb{I}/3$~\citep[P1 closure, but see][for other possible closure relations]{Thomas22,Hopkins22closure}. The interaction coefficient $\sigma_\kappa$ is a tensor decomposed into a $B$-field aligned component and a $B$-field perpendicular component, whose individual component values are proportional to $\nu/c^2 = (3\kappa)^{-1}$ (see details below), with $\nu$ the CR scattering rate, $c$ the speed of light, and $\kappa$ the CR diffusion coefficient. 
$\mathcal{L}_{\rm rad}$ and $\mathcal{L}_{\rm rad,c}$ represent thermal and CR radiative losses, respectively.
Finally, the gas momentum update by CRs, $-(\vec \nabla  P_{\rm c})^*$, contains a pressure gradient perpendicular to the $B$-field (corresponding to the CR Lorentz force $\vec{J}_{\rm c}\times \vec{B}$), $-\vec{\nabla}_\perp P_{\rm c}$, and a colinear $B$-field component proportional to the CR flux \citep{Zweibel13}
\begin{equation}
\label{eq:CR_gasmom}
 -(\vec\nabla P_{\rm c})^*= -\vec \nabla_\perp P_{\rm c}+ \vec b \sigma_\kappa \vec b.\left[ \vec{F}_{\rm c} - (\vec{u} + \vec{u}_{\rm s})(e_{\rm c} + P_{\rm c}) \right] \, ,
\end{equation}
where $\vec b=\vec B / \sqrt{\vec B . \vec B}$ is the normalised direction of the magnetic. This term is akin to the isotropic pressure gradient in the limit of steady-state flux (see below). It is worthwhile noting that the above equality relies on an assumption that cross field transport can be neglected. In practice, this assumption remains to be fully tested as field line random walk process is expected to control CR perpendicular transport to the mean field direction. 

The two-moment CR fluid equations for $e_{\rm c}$ and $\vec{F}_{\rm c}$ are derived by decomposing the CR distribution function $f(\vec{x}, \vec{p}, t) = f_0 + 3\mu f_1$, where $\mu = \vec{p}  . \vec{B} / |\vec{p}  . \vec{B}|$ is the pitch angle, into isotropic ($f_0$) and anisotropic ($f_1$) components. The Vlasov equation is similarly split and expanded to leading order in $u/c$ in $\mu$ (see~\citealp{Thomas19},~\citealp{Hopkins22closure}, and references therein).
Taking the CR kinetic energy moment (integrated over ${\rm d}\vec{p}$) of $f_0$, and the kinetic energy flux moment of $f_1$, assuming $\langle \mu \rangle \simeq 0$ (i.e.~near isotropic pitch angle distribution), using $p$-bin-centered approximations and a power-law slope in $p$ of $-4$ for the distribution function (i.e.~$f\propto p^{-4}$, e.g.~\citealp{Drury83}), and converting to lab-frame fluxes (neglecting non-leading order $c^{-2} \partial_t$ terms) leads to Eqs.~\ref{eq:ecr} and \ref{eq:fecr}.

Before moving forward, an important limit of this system of CR equations is established in the CR flux steady-state regime $c^{-2} \partial_t \vec{F}_{\rm c} \rightarrow 0$ (or $c \rightarrow \infty$), where the flux becomes 
\begin{align}
\label{eq:fcr_steady}
\vec{F}_{\rm c} = -\frac{1}{3\sigma_\kappa} \vec{\nabla} e_{\rm c} + (\vec{u} + \vec{u}_{\rm s})(e_{\rm c}+P_{\rm c}) \,,
\end{align}
and Eq.~\ref{eq:ecr} reduces to the familiar one-moment advection-diffusion equation for CR energy
\begin{align}
\label{eq:ecr_diff}
&\frac{\partial e_{\rm c}}{\partial t} + \vec{\nabla}  . \left[ (\vec{u} + \vec{u}_{\rm s})(e_{\rm c} + P_{\rm c}) \right] =  (\vec{u} + \vec{u}_{\rm s})  . \vec{\nabla} P_{\rm c}  \nonumber\\
&\qquad - \vec{\nabla} . \left(\frac{1}{3\sigma_\kappa} \vec{\nabla} e_{\rm c}\right) + \mathcal{L}_{\rm rad, c} \,,
\end{align}
assuming $\gamma_{\rm c} = 4/3$. In this regime, the right-hand side (RHS) of Eq.~\ref{eq:momentum} can be replaced by $-\vec{\nabla} P_{\rm c}$.

Coming back to Eqs~\ref{eq:ecr} and~\ref{eq:fecr}, these two-moment CR equations form a system of hyperbolic conservation laws with source terms that can be operator-split. Using a standard operator-splitting technique, the conservative part (i.e.~LHS terms) is solved with standard numerical flux solvers (see Section~\ref{section:conservative_step}) using a Courant-Friedrichs-Lewy (CFL) condition $\Delta t_{\rm c} \leq \Delta x / c$, while source (RHS) terms are treated separately (see Section~\ref{section:source_step}).
Since in most applications $c \gg u$, $u_{\rm A}$, and  $u_\kappa = \kappa |\vec{\nabla} e_{\rm c}| / e_{\rm c} \lesssim \kappa/\Delta x$ (the diffusion velocity), this two-moment formulation serves as a practical workaround to avoid the severe numerical limitations on time-stepping $\Delta t$ as a function of spatial resolution $\Delta x$ of diffusion-dominated ($\Delta t \propto \Delta x^2$) or streaming-dominated ($\Delta t \propto \Delta x^3$) evolution~\citep{Sharma10}.
To make the two-moment method tractable, we adopt the reduced speed-of-light approximation (as in radiative transfer; see \citealp{Gnedin01,Rosdahl13}) where $c$ in the $c^{-2} \partial_t \vec{F}_{\rm c}$ term is replaced by $\tilde{c} \ll c$ (but the $\sigma_\kappa = \nu/c^2$ term still uses the full speed of light). The chosen value of $\tilde{c}$ should be larger than any characteristic velocity of the plasma–CR fluid. Our experience has shown that $\tilde{c}$ on the order of or smaller than the maximum gas velocities and MHD wave speeds leads to negative CR energy densities, giving rise to catastrophic CR energy oscillations. Conversely, having diffusion or streaming velocities larger than $\tilde{c}$ does not lead to any numerical problems, but simply gives artificially low CR transport velocities.

Notably, the streaming term in the CR flux equation appears in addition to the fluid velocity, consistent with the derivation from $f_1$, unlike the formulation of~\JO{} (or~\citealp{Chan19}), where it is absorbed into the interaction coefficient. However, basic streaming tests show that the original form leads to numerical instabilities (see Appendix~\ref{appendix:streaming_vel_vs_damping}). For stability, we therefore follow \JO{} \citep[and][]{Chan19}, and split the interaction coefficient into $\sigma = \sigma_\kappa + \sigma_{\rm s}$. In the steady-state limit $\tilde c^{-2}\partial_t \vec{F}_{\rm c} \rightarrow 0$, both formulations are strictly equivalent.

The modified system for $e_{\rm c}$ and $\vec F_{\rm c}$ thus becomes (we drop $\mathcal{L}_{\rm rad, c}$ for clarity)
\begin{align}
&    \frac{\partial e_{\rm c}}{\partial t} + \vec{\nabla} . \vec{F}_{\rm c} = \vec{u} . \vec{\nabla} P_{\rm c}
- \vec{u}_{\rm s} . 3(\gamma_{\rm c} - 1) \sigma \left[ \vec{F}_{\rm c} - \vec{u} (e_{\rm c} + P_{\rm c}) \right] \, , \label{eq:ecr2} \\
&   \frac{1}{\tilde{c}^2} \frac{\partial \vec{F}_{\rm c}}{\partial t} + \frac{1}{3} \vec{\nabla} e_{\rm c} 
    = -\sigma \left[ \vec{F}_{\rm c} - \vec{u} (e_{\rm c} + P_{\rm c}) \right] \, , \label{eq:fecr2}
\end{align}
with
\begin{align}
&\frac{1}{3\sigma_\kappa} = \kappa \left[ f_{\parallel,\kappa} \vec{b} \otimes \vec{b} + (1 - f_{\parallel,\kappa}) \mathbb{I} \right] \,, \\
&\frac{1}{3\sigma_{\rm s}} = |\vec{u}_{\rm A}| (e_{\rm c} + P_{\rm c}) \left[ f_{\parallel,\rm s} \frac{\vec{b} \otimes \vec{b}}{|\vec{b}  . \vec{\nabla} e_{\rm c}|} + (1 - f_{\parallel,\rm s}) \frac{\mathbb{I}}{|\vec{\nabla} e_{\rm c}|} \right] \,,
\end{align}
decomposed into magnetic-field-aligned (anisotropic) and isotropic components, controlled by $0\leq f_{\parallel,\kappa} \leq 1$ and $0\leq f_{\parallel,\rm s} \leq 1$. The exact values of these two parameters depend on the microphysics of CR transport and on the turbulent properties of the gas (\citealp{Cho02,Sampson23}, e.g.~for super-Alfv\'enic turbulence CR transport becomes isotropic and, hence, $f_{\parallel,\kappa}=f_{\parallel,\rm s}=0$). In the rest of this paper we will use $f_{\parallel,\kappa}=1$ and $f_{\parallel,\rm s}=1$ as our default choice unless explicitly otherwise stated.

Similarly, replacing the RHS of the momentum equation (Eq.~\ref{eq:momentum}) with $-(\vec{\nabla} P_{\rm c})^*=-\vec{\nabla} P_{\rm c}$ provides a better approximation when $\tilde{c} \ll c$ than using Eq.~\ref{eq:CR_gasmom}, since deviations from equilibrium in the true equations are exaggerated in this regime. This substitution also yields better stability in strong CR shock tests (see Section~\ref{sec:tp_shock} for a 1D shock tube test) as opposed to using using Eq.~\ref{eq:CR_gasmom} (i.e. $-(\vec{\nabla} P_{\rm c})^*= \sigma ( \vec{F}_{\rm c} - (\vec{u} + \vec{u}_{\rm s})(e_{\rm c} + P_{\rm c}))$ in 1D). We therefore use $-(\vec{\nabla} P_{\rm c})^*=-\vec{\nabla} P_{\rm c}$ as our default gas momentum update by CRs in the two-moment formulation of CRs in~\ramses. The gas momentum equation is now
\begin{equation}
\label{eq:momentum2}
\frac{\partial (\rho \vec{u})}{\partial t} + \vec{\nabla}  . \left( \rho \vec{u} \otimes \vec{u} + P - \frac{\vec{B} \otimes \vec{B}}{4\pi} \right) = -\vec{\nabla} P_{\rm c} \,.
\end{equation}

We use the standard MHD solvers of \ramses~\citep{Fromang06,Teyssier06} to update the MHD variables, with the exception of the momentum equation (Eq.~\ref{eq:momentum2}), which includes the CR pressure gradient. After the MHD update, the two-moment CR equations are advanced using subcycling until one full MHD timestep is completed. During each CR sub-timestep, the gas momentum is updated with $-\vec{\nabla} P_{\rm c}$.

We apply operator splitting to solve the two CR equations numerically with an implicit-explicit (IMEX) time-integrator, with a conservative transport step (explicit integrator) and a source term step (implicit integrator). Each CR step is integrated over a CR timestep $\Delta t_{\rm c} \approx \Delta x / (3 \tilde{c})$ and subcycled relative to the MHD timestep of the simulation.

\subsection{Conservative step}
\label{section:conservative_step}

Equations~\ref{eq:ecr2} and~\ref{eq:fecr2} without the RHS~terms can be rewritten in a more compact form $\partial_tq+\nabla . f_q=0$, where 
\begin{equation}
q=\begin{pmatrix}e_{\rm c}\\F_{{\rm c},x}\\F_{{\rm c},y}\\F_{{\rm c},z}\end{pmatrix}\ {\rm and} \
f_q=\begin{pmatrix}
 F_{{\rm c},x} & F_{{\rm c},y} & F_{{\rm c},z} \\
 \frac{\tilde c^2}{3} e_{\rm c} & 0 & 0 \\
 0 & \frac{\tilde c^2}{3} e_{\rm c} & 0 \\
 0 & 0 & \frac{\tilde c^2}{3} e_{\rm c}
\end{pmatrix}
\label{eq:CR_cons_system}
\end{equation}
The corresponding numerical fluxes at the cell interface $i-1/2$ for direction $x$ (cells are centered in $x_{i-1}$, $x_i$, etc.) are obtained using the Lax-Friedrichs  (LF) flux function
\begin{equation}
    \label{eq:llf}
     \mathcal{F}_{i-1/2}^{\rm LF}=-\frac{\lambda}{2}(q_{\rm R}-q_{\rm L})+ \frac{f_{q,\rm L}+f_{q,\rm R}}{2}
\end{equation}
where $\lambda=\tilde c/\sqrt{3}$ is the signal speed, and the left (`L') and right quantities (`R') at the cell interface $i-1/2$ are derived from their cell-center values (resp. $i-1$ and $i$) with second order interpolation and the Van Leer slope limiter. As in \JO, we limit the signal speed if the CR diffusion over the timestep is expected to be small, by modifying it to 
\begin{equation}
\lambda= {\rm min} \left[  \frac{\ccr}{\sqrt{3}} R+\vert u\vert, \ \ \frac{\ccr}{\sqrt{3}} \right], \label{eq:sigspeed_mod}
\end{equation}
where 
\begin{equation}
 R=\sqrt{\frac{1-\exp(-\tau^2)}{\tau^2}}.
\end{equation}
We take the effective optical depth $\tau$ as the ratio between the diffusion time across the cell width, $t_{\rm diff} = \Delta x^{2} \sigma / 2$, and the local timestep length $\Delta t_{\rm c}$, giving
\begin{equation}
    \tau = f_\tau \frac{\Delta x^2 \sigma} {2 \Delta t_{\rm c}},  \label{eq:tau}
\end{equation}
where $f_\tau$ is a fudge factor equalling $1$ by default. This choice of reduction factor $R$ smoothly connects the wave speed $\lambda$ in the fast diffusion regime ($\tau\simeq0$, $R\simeq1$ and $\lambda=\tilde c/\sqrt{3}$). The modification of the wave speed (compared to $\tilde c/\sqrt{3}$) reduces the numerical diffusion of the scheme which becomes evident in static tests (see Appendix~\ref{appendix:static}). We note that \JO{} express the optical depth differently, as $\tau_{\rm JO}=f_\tau \Delta x \sigma \ccr$. However, since $\Delta t_{\rm c} \approx \Delta x / (3 \ccr)$, both variations give similar optical depths ($\tau\approx 3/2 \, \tau_{\rm JO}$). \JO{} also differ from us in using the Harten–Lax–van Leer inter-cell flux function, whereas we prefer using LF for stability (see \App{sec:stationary_test}).


Finally the CR energy density and CR energy flux are updated with this new numerical flux for each Cartesian direction (here only showing $x$):
\begin{equation}
    q_i^{n+1}= q_i^n - \frac{\Delta t_{\rm c}}{\Delta x}(\mathcal{F}^{\rm LF}_{i+1/2}-\mathcal{F}^{\rm LF}_{i-1/2})\, ,
\end{equation}
where $n$ and $n+1$ stand for values at time $t^n$ and $t^{n+1}=t^n+\Delta t_{\rm c}$.

\subsection{Note on the interpolation step}
\label{section:interpolation}

In classical second-order Godunov numerical schemes used to integrate the hydrodynamical Euler equations, each hydrodynamical primitive quantity, $\rho$, $ \vec u$, and $P$ is interpolated at cell interfaces independently from the others. This is not as straightforward for the system shown in \Eq{eq:CR_cons_system}, since the CR energy density and CR energy flux are linked by physics under the constraint $F_{\rm c} \le c e_{\rm c}$. With the reduced speed of light approximation, the condition is even more stringent, $F_{\rm c} \le \tilde{c} e_{\rm c}$. Therefore, we check if $\Fcr > \tilde{c} e_{\rm c}$ each time  an interpolation is needed (at cell interfaces, at fine-to-coarse interfaces, and when a cell is refined) or after the source step (see below), and if it does, reduce the magnitude of the flux vector so that $F_{\rm c} = \tilde{c} e_{\rm c}$. Note that similar issues arise with M1 radiative transfer methods \citep[e.g.][]{Audit02,Melon21}.

\subsection{Source step}
\label{section:source_step}

The second step is to solve for the non-conservative source terms on the RHS~of  equations \ref{eq:ecr2} and \ref{eq:fecr2}. 
This is handled through an implicit update
\begin{align}
&    \frac{e_{\rm c}^{n+1}-e_{\rm c}^{n}}{\Delta t} = -(\vec u+\vec u_{\rm s}).3(\gamma_{\rm c}-1)\sigma^n(\vec F_{\rm c}^{n+1}-\vec u\gamma e_{\rm c}^{n+1}) \, , \label{eq:src_ec} \\
&    \frac{1}{\tilde c^2}\frac{\vec F_{\rm c}^{n+1}-\vec F_{\rm c}^{n}}{\Delta t} = -\sigma^n\left(\vec F_{\rm c}^{n+1}-\vec u \gamma_{\rm c} e_{\rm c}^{n+1}\right)\, \label{eq:src_fc}, 
\end{align}
where $\vec \nabla P_{\rm c}^{n+1}$ has been replaced by $-3(\gamma_{\rm c}-1)\sigma^n(\vec F_{\rm c}^{n+1}-\vec u \gamma_{\rm c} e_{\rm c}^{n+1})$ using the steady-state flux approximation. This substitution avoids expressing $e_{\rm c}^{n+1}$ in terms of neighbouring cell values at the same time level, which would otherwise require a global mesh-wide update of the CR variables using iterative solvers such as those from the conjugate-gradient family (as an alternative, one could resort to a simpler but less accurate explicit update for the source term).

The interaction coefficient $\sigma$ is composed of an anisotropic and isotropic components that can be arbitrarily balanced. 
The first step is to rotate the flux $\vec F_{\rm c}$ along the $\vec B$ field. Once the flux is rotated along $\vec B$, the equations can be solved in this new coordinate system for the anisotropic and isotropic components before applying the final inverse rotation step to bring back the flux $\vec F_{\rm c}$  in the original Cartesian coordinate system of the box.

We note that for the streaming term entering in the interaction coefficient $\sigma$, its value can drop to zero (or $\kappa_{\rm s}$ reaches infinity) at extrema where $\vec \nabla e_{\rm c}=0$.
To mitigate numerical issues arising from this ill-conditioning, we always enforce a maximum value of the pseudo-diffusion coefficient of streaming in practice, i.e.~$(3\sigma_{\rm s})^{-1}={\rm min}(\kappa_{\rm s},\kappa_{\rm s,max})$, with $\kappa_{\rm s,max}$ a value that depends on the particular problem and unit system.

\subsection{CR subcycling and adaptive mesh refinement} \label{sec:subcycling}
Our explicit conservative CR step is subject to the Courant condition, with the CR timestep length determined at each adaptive mesh refinement (AMR) level by
\begin{equation} 
    \dtcr = f_{\rm c} \frac{\Delta x}{3 \ccr}, \label{eq_dtcr}
\end{equation}
where $f_{\rm c}<1$ is the Courant factor (set at $f_{\rm c}=0.8$ in all computations performed in this paper). Typically the CR timestep is much shorter than the MHD timestep $\dtmhd$ , which is determined by the maximum velocity and MHD wave speed at the given AMR level. Hence, performing one CR timestep after each MHD timestep, would force us to reduce the MHD timestep length $\dtmhd$ to $\dtcr$ and hence increase the computational cost. Our solution to this it to introduce CR subcycling, where we optionally perform up to $\Nsc$ CR timesteps for each MHD timestep. The constraint on the MHD timestep length on the given level is thus reduced from $\dtmhd \leq \dtcr$ to $\dtmhd \leq \Nsc \dtcr$. For a AMR given level, the sequence is thus: i) determine $\dtmhd$ from the maximum velocities and MHD wave speeds for all cells at the current level, and additionally $\dtmhd \leq \Nsc \dtcr$, ii) perform the MHD timestep, advecting the gas and magnetic field (including updates of neighbouring cells at the next coarser level), iii) perform up to $\Nsc$ CR timesteps, propagating CRs between cells, to fill in the MHD timestep.

Note that the this approach leads to the CR energy not being strictly conserved in the CR transport. This is due to the combination of the CR subcycling and AMR level subcycling, where two MHD timesteps are performed on a given level for one longer timestep on the next coarser level. This AMR subcycling is a huge advantage of computations in reducing the computational cost in the \ramses{}. Combined with CR subcycling, there can be many CR timesteps on a given AMR level before CR propagation is dealt with on the next coarser level, and in this case it is not feasible to update the coarser-level cell neighbours in each fine-level step. Hence we treat level boundaries with Dirichlet boundary conditions, fixing the values of CR density and flux at the coarser level neighbour but not updating it in sync when updating the fine-level cell due to the flux between the neighbouring cells at the fine and coarser level \citep{Commercon14}. In a nutshell, CRs flow out from (or into) the fine level boundary cell but not correspondingly into (or out from) its coarser level neighbour. When the CR propagation subcycling is later done on the coarser level, these neigbour cells to the next finer level are updated, but using the fixed CR variable states at the end of the fine-level subcycling (i.e.~out of sync). In practice, this leads to CR non-energy conservation, on the whole on the level of a few percent. Exactly the same kind of subcycling is done for the radiative transfer implementation in the \ramses{} code \citep[see][]{Rosdahl18}.  

\subsection{The variable speed of light} \label{sec:variable_c}
In reality, the propagation of CRs is limited by the speed of light $c$, and hence $c$ should set the maximum signal speed in our implementation. However, this would severely limit the timestep length (Eq.~\ref{eq_dtcr}) and make CR simulations unfeasibly expensive. The solution to this, proposed before by e.g.~\JO{}, is to replace $c$ with a reduced light speed\footnote{A similar philosophy, proposed by \cite{Gnedin01} is commonly used in radiative transfer. The idea is not completely the same though: in radiative transfer, the reduced speed of light changes the actual velocity of radiation, whereas in two-moment CR solvers, the reduced light speed only sets an upper limit to the diffusion, streaming, and advection speeds, and CRs may still propagate at lower speeds.} $\ccr \ll c$, the value of which is a fixed runtime parameter chosen by the user.

One must however be very cautious in the choice of $\ccr$: we find that choosing $\ccr$ lower than, or comparable to, the highest MHD wave speeds occurring in the simulation, negative CR energy densities appear. This is mostly an issue of the co-moving advection of CRs with the gas. If the CR signal speed is lower than that of the gas, the CR flux magnitude leads to a violation of the Courant condition. This sets a severe constraint when running simulations of real phenomena like galaxies, where the MHD wave speeds vary a lot over time and their maximum values are unknown a priori. The user then has to set a $\ccr$ for a given simulation in a very conservative way, making sure it is higher than any velocities that may occur throughout the simulation to be performed, and the resulting simulation becomes unnecessarily expensive due to this conservative choice.

We propose a method for addressing this problem, where we vary $\ccr$ over the runtime of the simulation, depending on the MHD wave speeds. Here, instead of a fixed $\ccr$, we fix a $\ccrmin$, which sets the \emph{minimum} CR wave speed (not to be confused with the MHD wave speed) throughout the simulation. If the maximum MHD wave speed $c_{\rm MHD}$ starts to approach and surpass this floor, $\ccr$ is dynamically adjusted so as to be significantly higher. By experience, we find that about $\ccr \approx 10 \, c_{\rm MHD}$ is sufficient to guarantee stability of the CR solver. In practice, we dynamically adjust $\ccr$ in every MHD timestep, and independently at each refinement level $\ell$ by requiring it to be higher or equal to $f_{\rm \ccr} c_{\rm MHD}$  at the given level, i.e. 
\begin{equation}
    \ccr_{\ell} = \max\left[ \ccrmin, \,  f_{\rm \ccr} \max_{c_{\rm MHD} \in \ell}(c_{\rm MHD})  \right], \label{eq:cvar}
\end{equation}
with $f_{\rm \ccr}$ an adjustable parameter equal to 10 by default. Optionally, we allow the user to set similar constraints for the variable $\ccr$ to exceed the CR diffusion speed
\begin{equation}
    \ccr_{\ell} = \max\left[ \ccrmin, \   f_{\rm \ccr} \max_{c_{\rm MHD} \in \ell}(c_{\rm MHD}), \   
    f_{\rm \ccr} \frac{\kappa}{\Delta x_{\ell}}, \  \right]. \label{eq:cvar_opt}
\end{equation}
This additional constraint on $\ccr$ is not as important as exceeding $c_{\rm MHD}$, as the latter prevents the code from crashing, whereas the former only prevent the CRs from propagating too slowly. The condition on $c_{\rm MHD}$ is in other words sufficient for the code to run smoothly, and the additional constraint may add to the computational cost. We present results of some of the tests later in the paper using the variable speed of light.

\section{Tests}\label{sec:tests}
We now present several classical CR tests taken from the literature. These are mostly taken from \JO. We compare to analytic expressions where possible, but otherwise to either the \athena{} implementation from \JO{} or the 1-moment implementation in \ramses{} from \cite{Dubois19}.

\subsection{Cosmic ray streaming in 1D} \label{sec:triangle_test}
We first consider a one-dimensional test of pure streaming, propagating an initially triangular profile of CRs, repeating the setup described in section 4.1.1.~of \JO{}. We disable gas advection and streaming-heating (i.e.~we set $\partial_t e_{\rm c}+ \vec \nabla.\vec F_{\rm c}=0$), and fix the Alfv\'en velocity to $\va=1\,\cms$. Diffusion is essentially disabled, with $\kappa=10^{-8} \ \kcrunits$. The initial CR energy profile is described by 
\begin{align}
    e_{\rm c}(x,t=0) = e_{{\rm c},1} - a|x|,
\end{align}
where $e_{{\rm c},1}=2 \, \Ecrunits$ is the CR energy density at the peak of the triangular profile at $x=0$ and $a=1 \, {\rm erg \, cm^{-4}}$ is the slope.  We use a domain width $L_{\rm box}=2 \, {\rm cm}$, with $x=0$ at the centre, and the boundaries are constructed so to maintain the unity slope of the CR energy profile.  We activate AMR, with a minimum refinement level of $\ell_{\rm min}=7$, corresponding to a coarse resolution of $128$ cells, and a maximum refinement level $\ell_{\rm max}=9$, corresponding to an effective maximum resolution of $512$ cells. Cells are refined if the relative difference in CR energy density between neighbours is larger than $10^{-3}$. Thus the whole box is refined to the maximum initially, but as a flat CR energy profile develops at the centre, it is de-refined, and hence the refinement level changes around the moving knees where the CR energy profile turns from sloped to flat. We do not subcycle the CR solver on individual refinement levels, i.e.~the CR solver is strictly energy-conserving. We fix $\ccr=100 \ {\rm cm \ s^{-1}}$ in this test.

\begin{figure}
  \centering
  \includegraphics[width=0.37\textwidth]
    {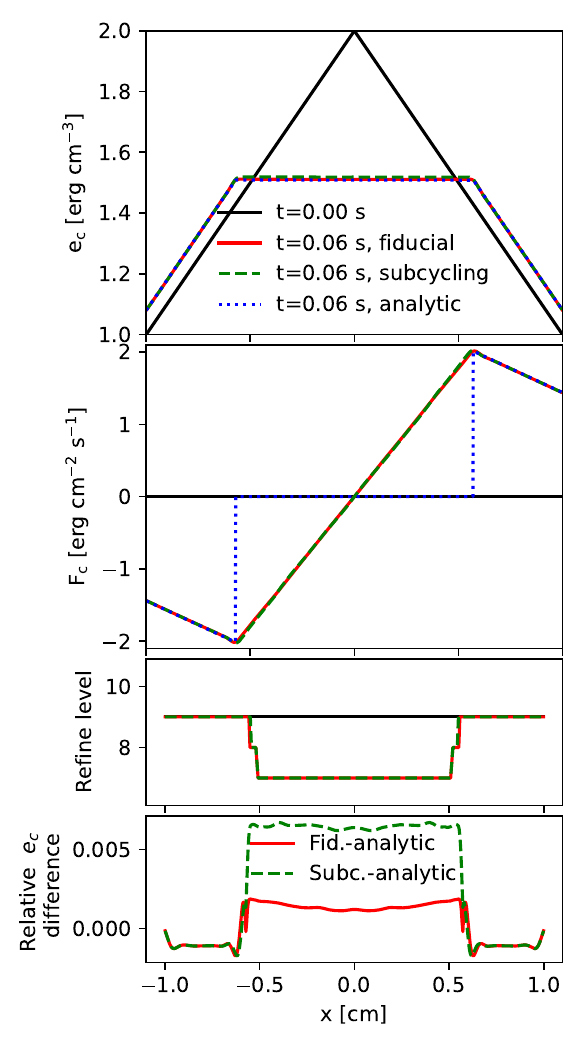}
  \caption
  {\label{fig:JO411_tri} CR streaming test in 1D with an initially triangular profile for the CR energy, as described in \Sec{sec:triangle_test}. From top to bottom, the panels show the CR energy density, the CR flux, the refinement level, and the relative difference between the numerical results and the analytic expectation. The black curves show the initial conditions, while the red solid curves show the result from the fiducial run at $t=0.06\,\rm s$, and the green dashed curves show the result from an analogue run with CR subcycling turned on, at the same time. The dotted blue curve in the top panel shows the analytic profile for the CR energy at $t=0.06\,\rm s$, and in the panel below it shows the steady state solution computed from the analytic CR energy density. The bottom panel shows that the fiducial run agrees well with the analytic expectation, with a maximum overshoot of about $0.2$ percent in $\Ecr$, while activating subcycling increases the overshoot to about of $0.6$ percent.}
\end{figure}

\JO{} derive an analytic solution for how the triangular CR energy profile develops with time. The triangle sides shift towards the edges of the box at a velocity of $4/3 \va$ and the top of the triangle flattens (as a result of $\kappa_{\rm s}\propto 1/\nabla e_{\rm c}$) so as to conserve energy, with the locations $\pm x_{\rm m}$ marking where the profile turns from flat to sloped. The evolving profile is thus described as
\begin{align}
    \Ecr(x,t) = 
    \begin{cases}
        e_{{\rm c},1} - a|x| + \frac{4}{3}a \va t,
        & \text{if } |x|> x_{\rm m}\\
        e_{{\rm c},1} - a x_{\rm m} + \frac{4}{3}a \va t,
        & \text{otherwise.}
    \end{cases}
\end{align}
The locations $\pm x_{\rm m}$ are derived from energy conservation as 
\begin{align}
x_{\rm m} = \sqrt{\frac{8}{3} \va t \left( \frac{e_{{\rm c},1}}{a} + \frac{2}{3}\va t \right)}\, .
\end{align}

We show the results of this streaming test in \Fig{fig:JO411_tri}. The top panel shows the CR energy density, with the initial profile in black, the numerical solution at $t=0.06 \, {\rm s}$ in red, and the analytic solution in dotted blue. The second panel from top shows the CR flux in the numerical solution at the initial and final times, as well as the steady state flux solution\footnote{The steady state value is $\vec F_{\rm c}=-\gamma_{\rm c}\vec u_{\rm A} e_{\rm c}{\rm sign}(\vec b.\vec \nabla e_{\rm c})$ where $\vec \nabla e_{\rm c}\ne 0$, and $\vec F_{\rm c}=0$ otherwise (because streaming is modeled as a pseudo-diffusion term).} (in dotted blue), calculated from the analytic solution for $e_{\rm c}$. The third panel shows the refinement levels, illustrating how the flat part of the CR energy profile is de-refined. The bottom panel shows (in solid red) the relative difference between the numerical solution and the analytic one.
The numerical solution is in good agreement with the analytic one, with an overshoot in $\Ecr$ of about $0.2$ percent in the flat part of the profile. We note that with a ten times higher $\ccr$ (not shown), the small disagreement with the analytic solution essentially disappears, except for small remaining differences of about $0.2$ percent at the 'knees' where the shape of the profile changes, which are there regardless of whether or not AMR is used.

We also demonstrate in \Fig{fig:JO411_tri} how the code performs with CR subcycling, with 10 CR steps in sequence on a given level, while other levels are frozen. The results for this test are shown in dashed green curves. As discussed in \Sec{sec:subcycling}, subcycling the CR steps increases the speed of the code (if gas advection is activated, though this is not the case here), but at the sacrifice that strict energy conservation is no longer maintained for CRs propagating over boundaries between refinement levels. Indeed, the green dashed curve in the bottom panel of \Fig{fig:JO411_tri}, comparing the relative difference between this subcycling test and the analytic solution, shows that in this case the numerical solution overshoots the analytic one a bit more here, or by about $0.6$ percent.

\subsection{Diffusion and advection in 1D} \label{sec:diffusion_test}
Here we perform a 1D test of combined CR diffusion and advection. In a domain with dimensions $[-1,1]\,\rm cm$, we set up a Gaussian CR energy profile at the centre:
\begin{align}
    \Ecr = e_{\rm c,0} \exp \left( -a x^2\right),
\end{align}
with $e_{\rm c,0}=1\, \Ecrunits$ and $a=40 \, {\rm cm^{-2}}$. We allow the CRs to diffuse, with a diffusion coefficient $\kappa = 0.1 \, \kcrunits$, and advect, with a gas velocity $\vgas=1 \,\cms$. Streaming is turned off and the gas is static, to prevent the gas from reacting to the  CR pressure, which would complicate the comparison to the analytic solution. We use periodic boundaries and $\ccr=10^2 \, {\rm cm \, s^{-1}}$. The coarse resolution level is set at $\ell_{\rm min}=7$, corresponding to 128 cells, and the code is allowed to refine up to level $\ell_{\rm max}=9$, corresponding to an effective resolution of 512 cells, when the relative difference in $\Ecr$ between neighbouring cells is larger than $10^{-2}$.

\begin{figure}
  \centering
  {\includegraphics[width=0.37\textwidth]
    {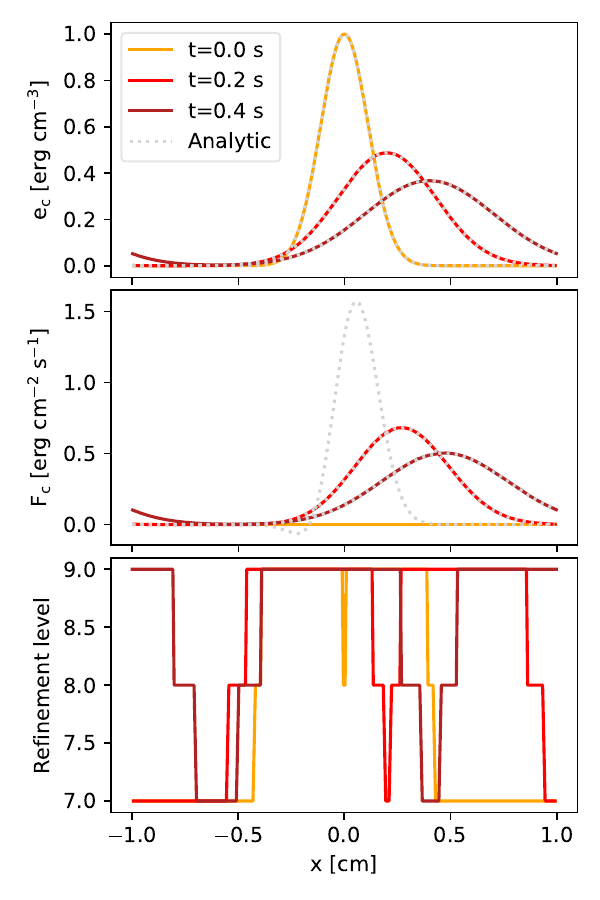}}
  \caption
  {\label{fig:test_JO414}CR advection-diffusion test in 1D as described in \Sec{sec:diffusion_test}. The top panel shows the CR energy density profile at three different times, the middle panel shows the same for the CR flux, and the bottom panel shows the refinement levels. The solid curves show the numerical test results. The grey dotted curves show the analytic solutions at the same times, in very good agreement with the numerical results.}
\end{figure}

The evolution of $\Ecr$ with time is described by the analytic solution to the diffusion equation, shifted due to the gas velocity:
\begin{align}
    e_{\rm c}(x,t) = \frac{e_{\rm c,0}}{\sqrt{1+4 a \kcr t}} \exp{\left(  \frac{-a(x-\vgas t)^2}{1+4 a \kcr t} \right)}.
\end{align}
The analytic steady-state solution for the CR flux is then 
\begin{align}
    \Fcr(x,t)&= \left[ \frac{2a\kappa (x-ut)}{1+4 a \kcr t}+u\gamma_c \right]\Ecr(x,t).
\end{align}
We compare in the top panel of \Fig{fig:test_JO414} the evolution of the CR energy density profile to the analytic solution at $t=0$, 0.2, and 0.4\,s. The numerical results are indistinguishable from the analytic solution, except at $0.4$\,s close to the left boundary, as we do not assume periodic boundaries in the analytic model. The middle panel shows the CR flux, with the steady-state analytic solution over-plotted. The numerical CR flux agrees very well with the analytic solution, except for the periodic boundary and time zero, where we impose a zero flux in the initial conditions (i.e. not matching the steady state). We note for this test that it performs equally well with zero gas velocity, i.e.~in the pure-diffusion case, and that subcycling the CR steps 10 times on each refinement level has a very similar effect as described in the streaming test in \Sec{sec:triangle_test}, i.e.~leading to few tenths of a percent overshoot in the peak CR energy density. 

\subsection{2D anisotropic diffusion in a looped magnetic field} \label{sec:2d_diffusion}

\begin{figure}
  \centering
  \subfloat
  {\includegraphics[width=0.45\textwidth]
    {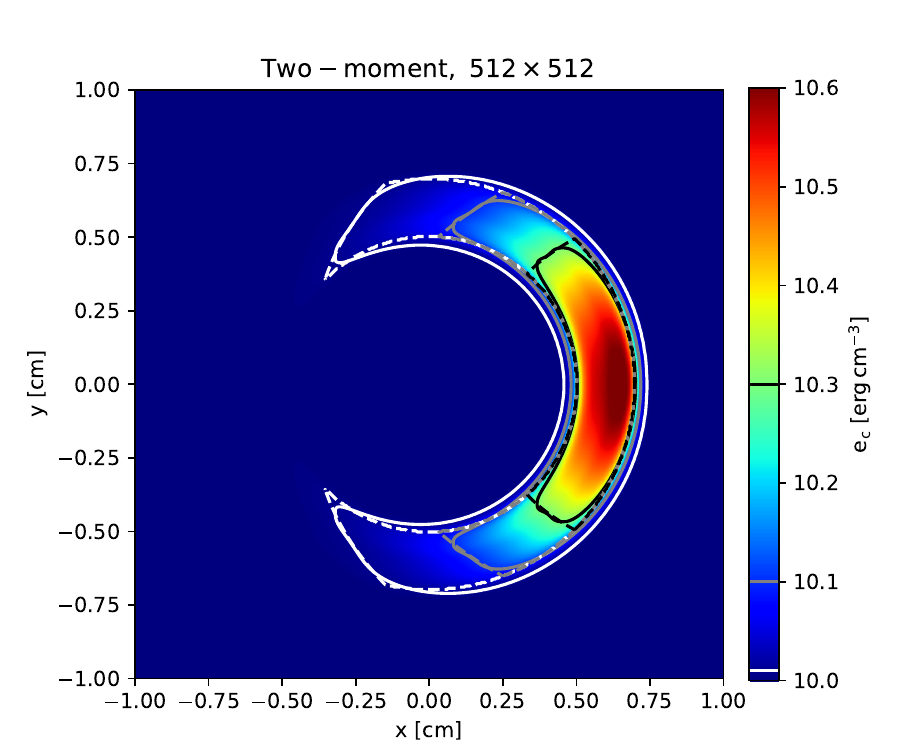}}
  \vspace{-9.2mm}
  \subfloat
  {\includegraphics[width=0.45\textwidth]
    {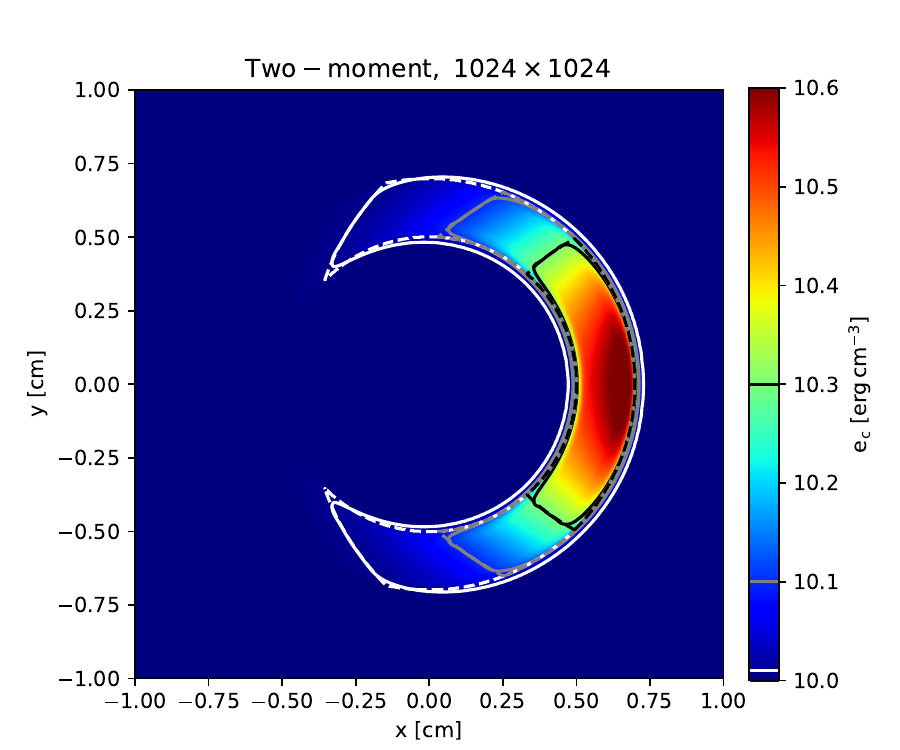}}
  \vspace{-9.2mm}
  \subfloat
  {\includegraphics[width=0.45\textwidth]
    {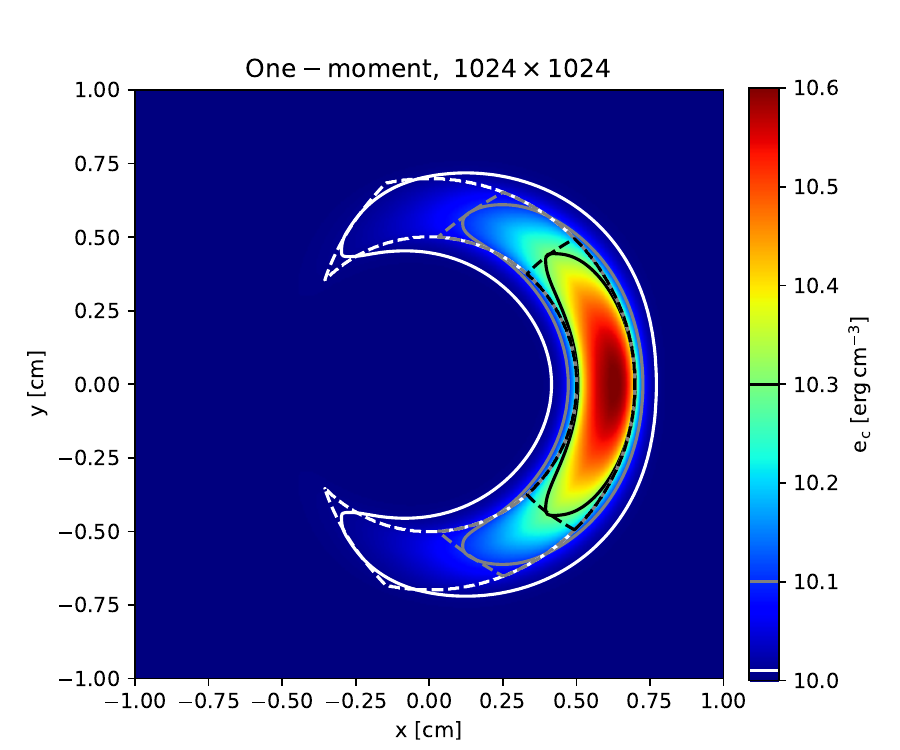}}
  \caption
  {\label{fig:test_donut} 2D anisotropic diffusion test, described in \Sec{sec:2d_diffusion}. We show maps $\Ecr$ at $t=0.26\,\rm s$ in a circular magnetic field. In each panel, three solid contours mark where $\Ecr=10.01$, $10.1$, and $10.3 \, {\rm erg \, cm^{-3}}$ with \ramses{}, and three dashed contours mark the same values for the analytic solution at the same time. The top panel shows the result for our two-moment method and a resolution of $512^2$ cells ($\ell=9$), while the middle panel shows the same with one-level higher resolution, i.e. $1024^2$ cells ($\ell=10$). The bottom panel shows an equivalent run, also with a resolution of $1024^2$, with the one-moment solver.}
\end{figure}

\begin{figure*}
  \centering
  \subfloat
  {\includegraphics[width=0.2993\textwidth]
    {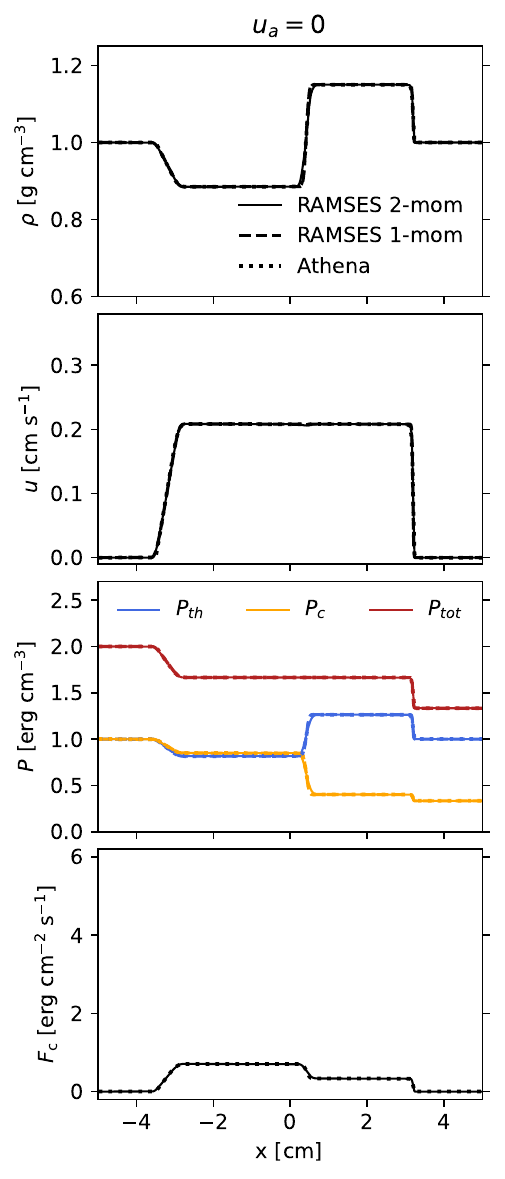}}
  \hspace{-12.0mm}
  \subfloat
  {\includegraphics[width=0.3\textwidth]
    {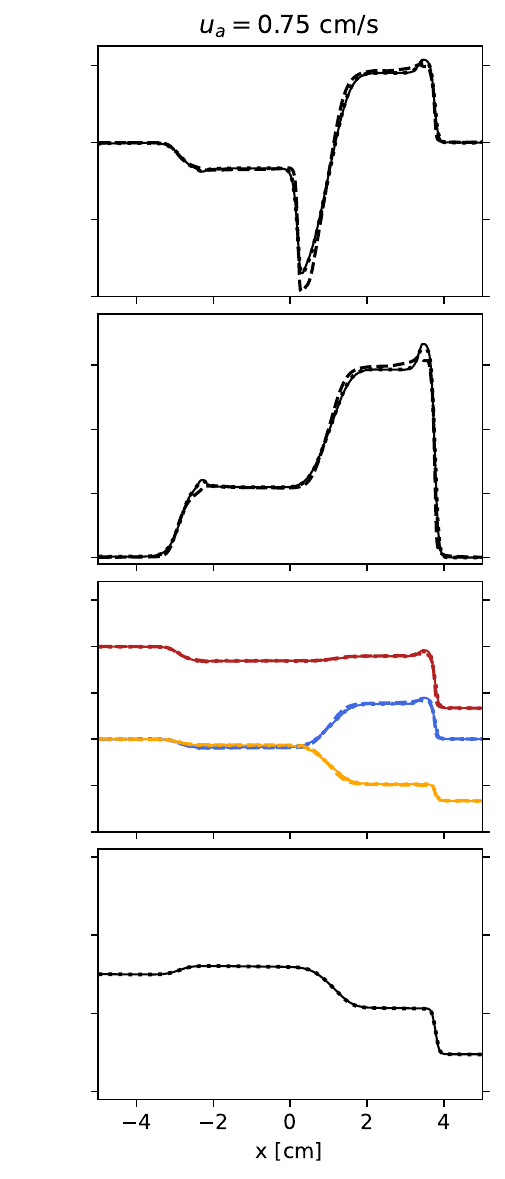}}
  \hspace{-12.0mm}
  \subfloat
  {\includegraphics[width=0.3\textwidth]
    {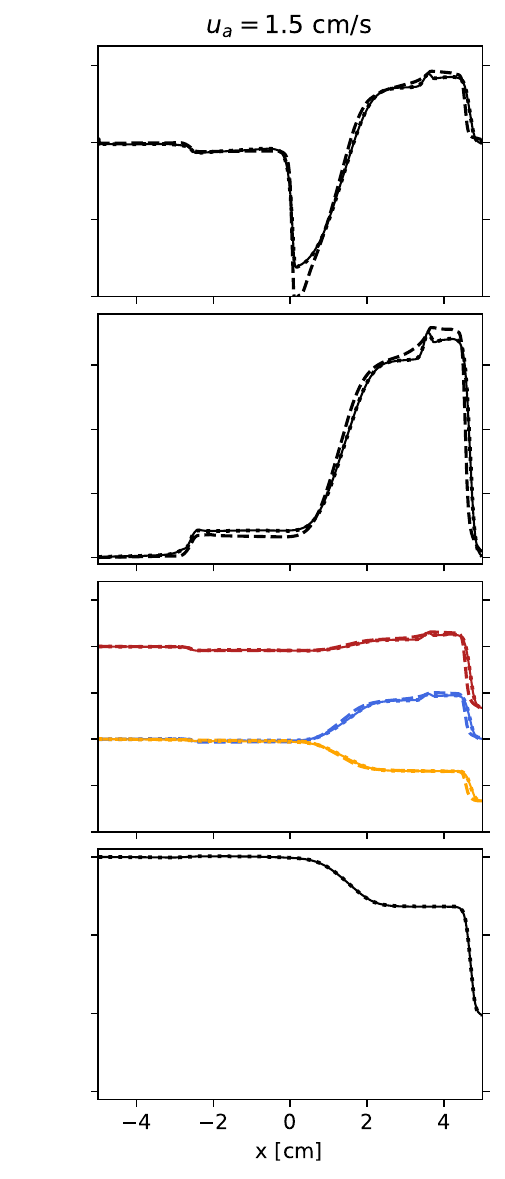}}
  \caption
  {\label{fig:test_shock} Shock tube test, described in \Sec{sec:tp_shock}. We show results run with the two-moment CR implementation in \ramses{} (solid curves), compared with the one-moment implementation in the same code (dashed curves), and \athena{} (dotted curves), for different values of the fixed Alfv\'en speed, $\va=0, 0.75$, and $1.5\,\cms$, from left to right. The panels show profiles at $t=2\,\rm s$ of, from top to bottom, gas density, gas velocity, pressures (of gas, CRs, and their sum), and CR flux.}
\end{figure*}

We now test the classic anisotropic diffusion of CRs along a circular magnetic field in two dimensions. We set up a square 2\,cm wide domain with a circular magnetic field around the centre at $(x,y)=(0,0)$ and an over-density in the CR energy density along an annular sector,
\begin{align}
    \Ecr(x,y) = 
    \begin{cases}
        e_{\rm c,1}
        & \text{if } r \in [0.5, 0.7] \text{ and } |\phi| < \pi/12 \\
        e_{\rm c,0}
        & \text{otherwise,}
    \end{cases}
\end{align}
where $r=\sqrt{x^2+y^2}$, $\phi$ is the azimuthal angle from the $x$-axis, $e_{\rm c,0}=10 \, {\rm erg \, cm^{-3}}$, and $e_{\rm c,1}=12 \, {\rm erg \, cm^{-3}}$. $\Fcr$ is initialised to zero. We apply a diffusion coefficient of $\kcr=1/3 \, \kcrunits$ parallel to the magnetic field, and a factor $10^6$ smaller perpendicular to it. MHD is turned off in this test (i.e.~the gas and magnetic fields are static) and so is the streaming of CRs.

The top panel of \Fig{fig:test_donut} shows the diffusion of $\Ecr$ at $t=0.26\,\rm s$ with a resolution of $512^2$ cells. The CR energy density propagates in the magnetic-field loop without creating any inappropriate negative energy densities. We have marked solid contours at selected values for $\Ecr$ as indicated in the color bar. They can be compared to the dashed contours, which mark the same values for the analytic solution for the circular diffusion \citep[see \JO{} and][]{Pakmor16}:
\begin{align}
    e_{\rm c,ana}(x,y) &= e_{\rm c,0} \\ 
                       &+\left(e_{\rm c,1} - e_{\rm c,0} \right) \nonumber \\
                       & \times
                  \left( \operatorname{erfc}\left[ \left(\phi - \frac{\pi}{12} \right) \frac{r}{D}\right]
                  - \operatorname{erfc}\left[ \left(\phi + \frac{\pi}{12} \right) \frac{r}{D}\right] \right).
                  \nonumber
\end{align}
Here $\operatorname{erfc}$ is the complementary error function and $D=\sqrt{4t\kcr}$. Based on the contours, the agreement with the analytic solution is quite good, though with some diffusion perpendicular to the magnetic field. We show in the middle panel an identical run with one level higher resolution, i.e.~$1024^2$ cells ($\ell=10$). Here we inch closer to the analytic solution, compared to the lower-resolution equivalent in the top panel. The bottom panel finally shows the same setup, also at a resolution of $1024^2$ cells, performed with the one-moment CR solver of \cite{Dubois16} in \ramses, and setting a perpendicular diffusion coefficient $10^{12}$ smaller than the parallel one. Here we get significantly more numerical diffusion perpendicular to the magnetic field, compared to the two-moment method. 

\subsection{Shock tube with and without streaming} \label{sec:tp_shock}

\begin{figure}
  \centering
  {\includegraphics[width=0.3\textwidth]
    {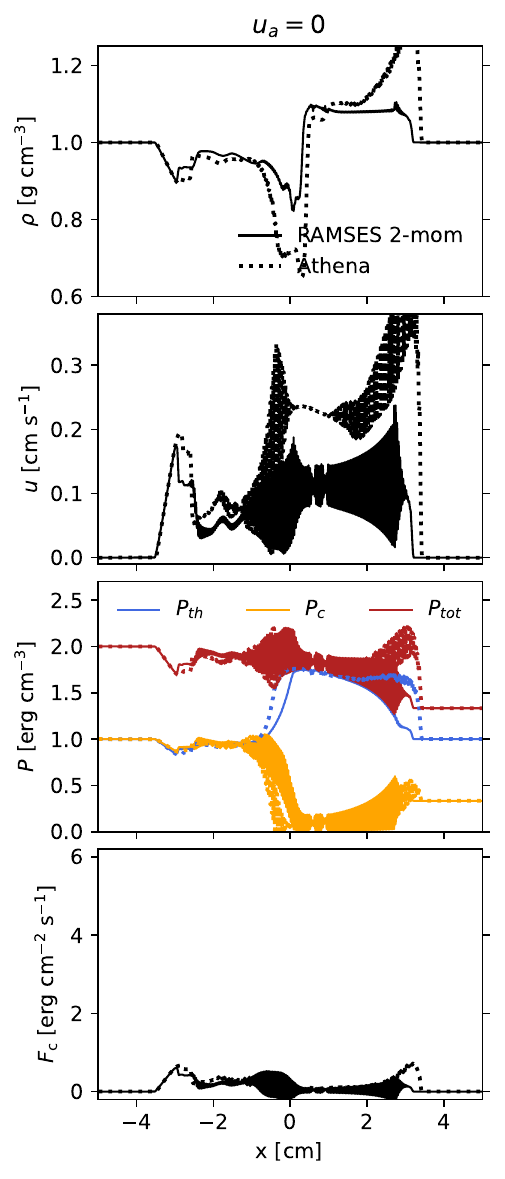}}
  \caption
  {\label{fig:test_shock_unstable} Same shock test as in the $\va=0$ case in \Fig{fig:test_shock}, at $t=2\,\rm s$, but demonstrating how numerical instabilities can easily arise if care is not taken. In the case of \ramses, we have run with the HLLE inter-cell flux function and CR to gas momentum injection using the approach of \JO. In the case of \athena{}, we have run with the cell optical depth, $\tau$, manually reduced by a factor 2 compared to the \JO{} formulation.}
\end{figure}

We repeat here the one-dimensional shock test described in section 4.2 of \cite{Thomas21}. We set up a domain $x \in [-5, 5]\,\rm cm$ with an initial homogeneous gas density $\rho=1 \ {\rm g \ cm^{-3}}$, gas pressure $\Pgas=1 \ {\rm erg \ cm^{-3}}$, magnetic field $\vecB=(10^{-13},0,0)$ G, and zero gas velocity. We initialise a CR pressure imbalance between the two sides of the domain:
\begin{equation}
    \Pcr=
    \begin{cases}
        1 \ {\rm erg \ cm^{-3}}
        & \text{where} \ x < 0  \\
        \frac{1}{3} \ {\rm erg \ cm^{-3}}
        & \text{elsewhere,}
    \end{cases}
\end{equation}
with $\Fcr=\va(\Ecr+\Pcr)$ everywhere. The resolution is fixed at 512 cells, i.e.~we do not use AMR, we fix the maximum CR propagation speed at $\ccr=100\,\cms$, and we do not use CR subcycling in this test. We run three experiments, with fixed Alfv\'en velocities $\va=0, 0.75, $ and $1.5\,\cms$, in all cases using a diffusion coefficient of $\kappa=1/300 \, {\rm cm^2 \, s^{-1}}$. We run with streaming losses and transport (when $\va\ne0$) and the MHD are activated so as to capture the interaction of the CRs and gas. 

We consider in \Fig{fig:test_shock} the test results for these three cases, comparing the results at $t=2\,\rm s$ of our new two-moment implementation in \ramses{} with those of the one-moment implementation in the same code, described in \cite{Dubois16} for anisotropic diffusion and~\cite{Dubois19} for streaming\footnote{In short, the one-moment implementation employs an implicit method using a minmod limiter on the (cell face-oriented) transverse flux with BiCGSTAB method to solve for both diffusion and streaming, treating streaming as a pseudo-diffusion term.}, and with those of \athena{} as described in \JO. For \athena{}, we have run with the same setup and resolution, using the 'Van-Leer 2' integrator, and second-order for both gas and CRs. For the $\va=0$ case, we use an optical depth fudge-factor in \athena{} of $f_{\tau}=0.1$ (see Eq.~\ref{eq:tau} and discussion below it) to prevent numerical instabilities. We show, from top to bottom, profiles of gas density, gas velocity, pressure (of gas, CRs, and combined), and CR flux (only shown for the two-moment case, since the one-moment method does not evolve the flux). The columns, from left to right, show the tests with $\va=0$, $0.75$, and $1.5\,\cms$, respectively.  The CR pressure difference (yellow curves) generates a shock propagating rightwards, with the detailed shock structure and extent depending on the value of $\va$. Overall the 1-moment and 2-moment methods give very similar results for all three values of $\va$, and those of \athena{} are almost identical to the 1-moment run in \ramses, as expected. We note that these results are well converged with respect to variations in $\ccr$ and resolution. Our results are also very similar to those reported in \cite[not shown]{Thomas21}, using their two-moment CR implementation in the \arepo{} code~\citep{Springel10}, in terms of the general shapes of profiles and the locations of discontinuities at $t=2\,\rm s$. There is a notable difference though in the detailed shapes of the profile for nonzero $\va$: \ramses{} (and \athena) exhibit somewhat smoother shock transitions and notable `bumps' close to both the forward and reverse shocks, which are absent in \arepo. The reason for these differences are not clear to us.  We note that although we do not use CR subcycling here, we have replicated the tests with 10 subcycling CR steps per MHD timestep, and it has indiscernible effect on the results.

The leftmost column of panels in \Fig{fig:test_shock} represents the case where $\va=0$. This low-diffusion case falls in a $\tau \gg 1$ regime which is prone to numerical instabilities\footnote{Using Eq.~\ref{eq:tau} with our setup parameters (and replacing $\Delta t = \Delta x/\ccr$) gives $\tau \approx 300$.}. In general, including in \Fig{fig:test_shock}, we avoid those instabilities in \ramses{} by fiducially i) utilising the LF inter-cell flux function, introducing some numerical diffusion, and ii) by injecting momentum from CRs to gas by directly using the gradient of CR pressure, i.e.~$\partial_t (\rho \vec{u}) = -\vec{\nabla} \Pcr$.  Severe instabilities appear, however, if we instead follow the implementation of \JO{}, by a) using the less diffusive HLLE inter-cell flux function and b) using  $\partial_t (\rho \vec{u}) = -\sigma ( \vec{F}_{\rm c} - \vec{u} (e_{\rm c} + P_{\rm c}))$. For demonstration, we show in \Fig{fig:test_shock_unstable} these instabilities in the non-streaming shock test from the left column of \Fig{fig:test_shock}, using variations a) and b) above. We also show the same setup run with \athena{}, producing qualitatively similar instabilities\footnote{We use here $f_\tau=1.0$ in \ramses{} but  $f_\tau=0.5$ in \athena{}, as using a full $f_\tau=1$ leads to tiny timestep lengths with \athena{} and the test cannot reach $t=2\,\rm s$}. Either of our modifications i) (LF inter-cell flux) or ii) (CR pressure gradient momentum injection) is independently sufficient to mostly avoid the instabilities shown, though the use of the CR pressure gradient alone (with the HLL inter-cell flux) still shows small oscillations in the gas density in the right side of the domain (not shown).

\subsection{Interactions of cosmic rays with a gas cloud}\label{sec:cloud}
\begin{figure*}
  \centering
  \subfloat
  {\includegraphics[width=0.37\textwidth]
    {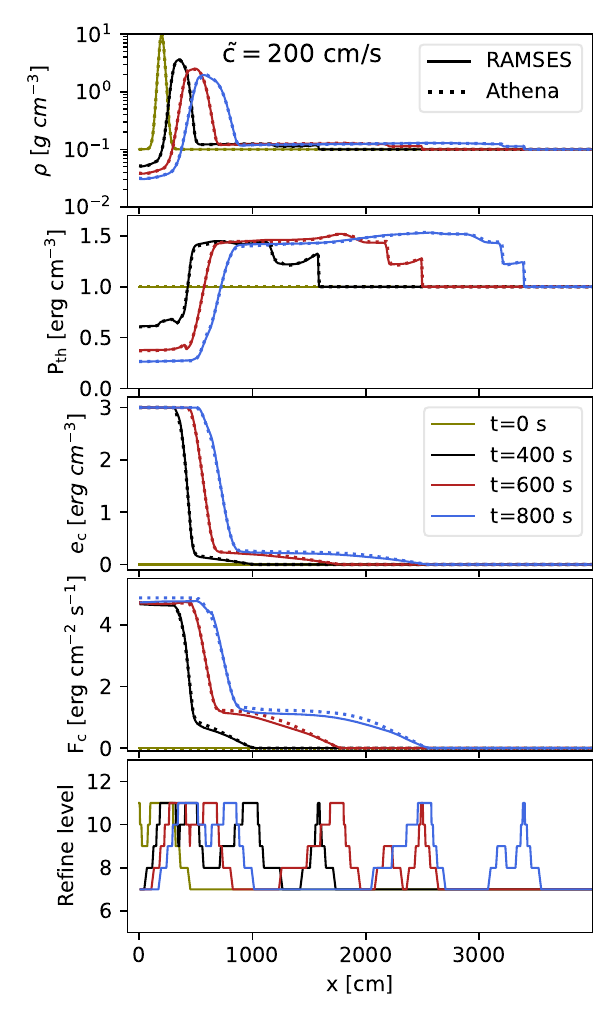}}
  \hspace{-15.5mm}
  \subfloat
  {\includegraphics[width=0.37\textwidth]
    {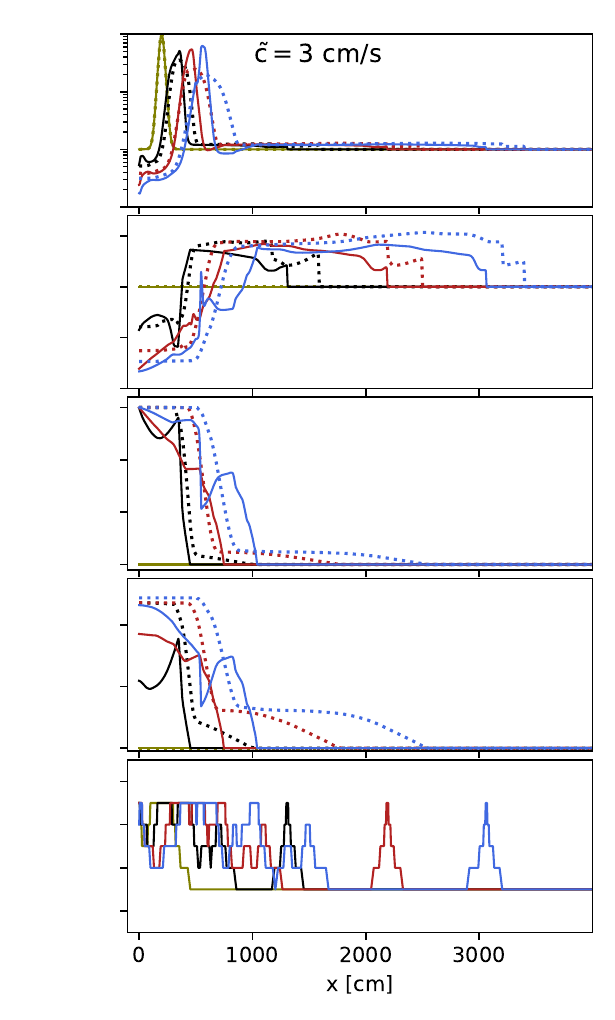}}
  \hspace{-15.5mm}
  \subfloat
  {\includegraphics[width=0.37\textwidth]
    {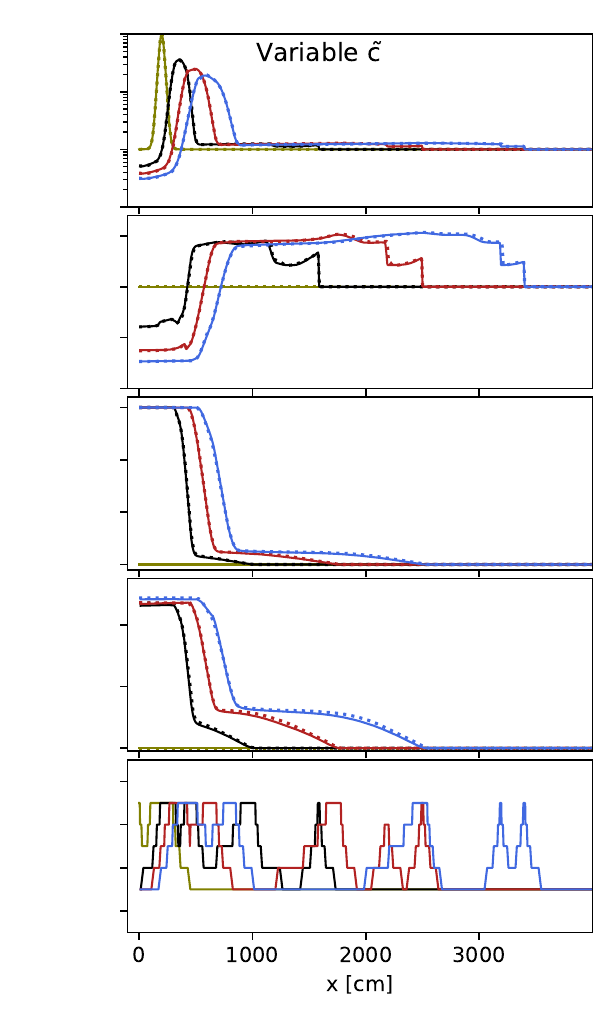}}  \caption
  {\label{fig:test_JO424} CRs-cloud interaction test, described in \Sec{sec:cloud}. Each column of plots shows profiles of gas density, gas pressure, CR energy density, CR flux, and refinement level, from top to bottom, at different times as indicated in the legend, for \ramses, in solid, and \athena, in dotted curves. The leftmost column shows results with a fixed $\ccr=200 \, \cms$, both in \ramses{} and \athena. The middle column is the same except we set $\ccr=3 \, \cms$ in \ramses{} (but keep the fiducial $\ccr=200 \, \cms$ in \athena{} for reference). The rightmost column is with a variable $\ccr$ in \ramses{}.}
\end{figure*}

We now turn to a reproduction of the test described in section 4.1.4 of \JO{}. This is a one-dimensional test with full coupling between CRs and gas, mimicking the CRs streaming through and beyond a dense cloud. With a homogeneous magnetic field in the simulated volume, the Alfv\'en velocity decreases within the high-density cloud, and the cloud creates a bottleneck for the streaming of CRs, while at the same time being pushed by them.

We initialise a domain $x \in [0, 4000]\,\rm cm$ with a gas density concentration close to the origin:
\begin{align}
    \rho(x) &= \rho_{\rm h} + (\rho_{\rm c} - \rho_{\rm h}) \left[ 1 + \tanh{\left( \frac{x-x_0}{L_{\rm c}} \right) }\right] \nonumber \\
    & \times \left[ 1 + \tanh{\left( \frac{x_0-x}{L_{\rm c}} \right) }\right], 
\end{align}
with ambient gas density $\rho_{\rm h}=0.1\,\rm g\,cm^{-3}$, cloud peak density $\rho_{\rm c}=10 \,\rm g\,cm^{-3}$, cloud center at $x_0=200\,\rm cm$, and cloud width $L_{\rm c}=25\,\rm cm$. Other quantities are initialised to uniform values: a gas pressure of $P_{\rm gas}=1 \,\rm erg\,cm^{-3}$, zero gas velocity, magnetic field magnitude of $\vecB=(\sqrt{4\pi},0,0) \,\rm G$, which, thus, stays fixed with time, CR energy density of $\Ecr=10^{-6}\,\rm erg\,cm^{-3}$, and zero CR flux. The Alv\'en velocities are $3.2$ and $0.32\,\cms$ in the ambient gas and at the peak cloud density, respectively.

We fix a small diffusion coefficient of $\kappa=10^{-6} \, {\rm cm^{2}\,s^{-1}}$, but activate streaming (transport and loss terms) for the CRs. The left boundary, corresponding to $x=0$, has an imposed $\Ecr=3\,\rm erg\,cm^{-3}$, leading to a flow of CRs into the domain from the left side. The boundary is reflective in gas velocity and an outflow boundary for other quantities. We use pure outflow conditions for the right boundary. We allow 10 CR subcycling steps in \ramses{} over the MHD timestep and we use AMR going from level $\ell_{\rm min}=7$, corresponding to a coarse resolution of 128 cells, to level $\ell_{\rm max}=11$, corresponding to an effective fine resolution of 2048 cells. We refine if the relative difference between neighbouring cells is larger than 5 percent for gas density, gas pressure, gas velocity, or CR energy density. For comparison, we run the same setup with \athena{}, except here we use a fixed resolution of $2048$ cells. We use a fiducial fixed $\ccr=200 \, \cms$ in both \ramses{} and \athena{} (as done in \JO), but perform additional \ramses{} runs, one with $\ccr=3 \, \cms$ and another where we activate the variable $\ccr$ approach, with a \emph{floor} value of $\ccr=3 \, \cms$, activated to exceed $c_{\rm MHD}$ but not the CR diffusion speed, which is anyway orders of magnitude lower than $c_{\rm MHD}$.

\begin{figure*}
  \centering
  \subfloat
  {\includegraphics[width=0.99\textwidth]
    {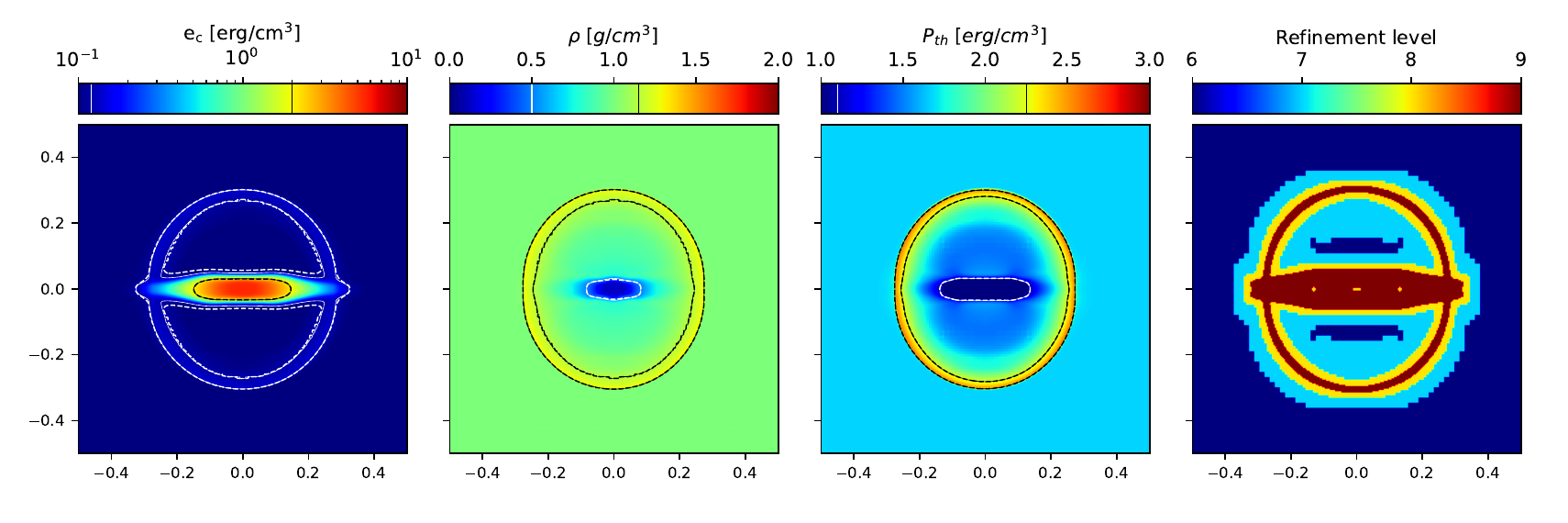}}
  \vspace{-5mm}
  \hspace{-3mm}
  \subfloat
  {\includegraphics[width=0.99\textwidth]
    {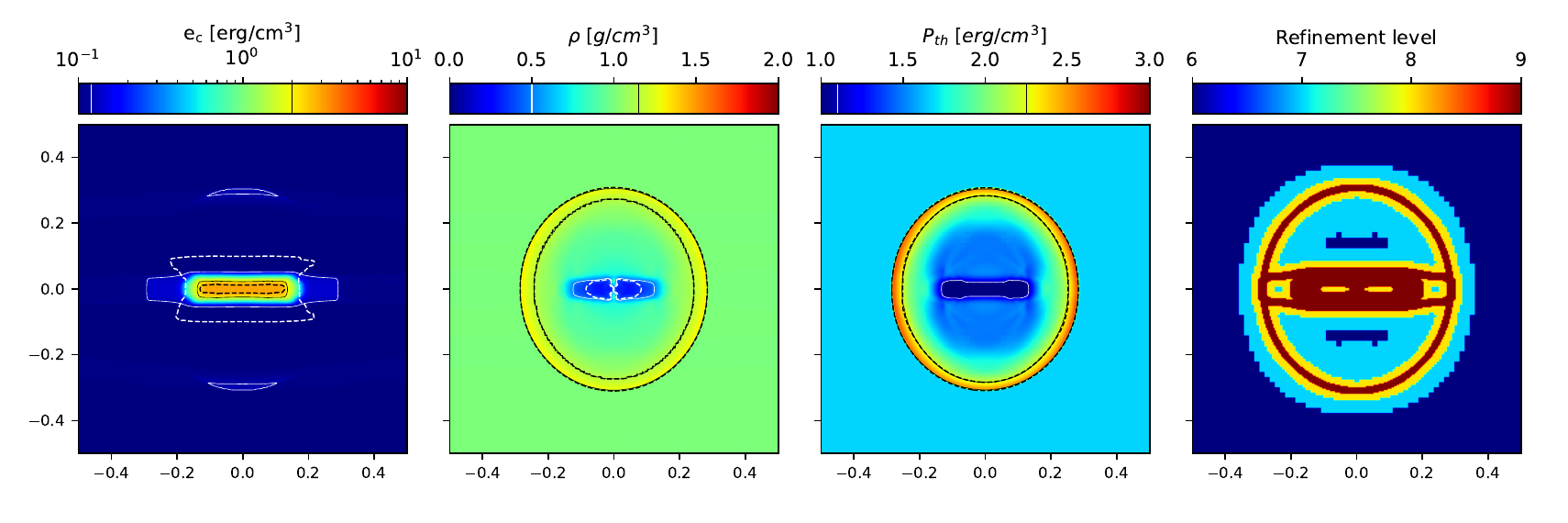}}
  \caption
  {\label{fig:test_JO423} CR-driven blast test, described in \Sec{sec:2dblast}. All panels show results at $t=0.1\,\rm s$. The upper row is with pure diffusion and the lower row is with pure streaming. From left to right, the panels show CR energy density, gas density, gas thermal pressure, and refinement level. For the gas and CR quantities, we over-plot contours at arbitrarily selected values, as indicated in the color-bars, with solid representing the current two-moment method in the maps, and dashed representing equivalent runs with the one-moment CR integration method. The two methods agree very well for the diffusion case, but not as well with pure streaming.}
\end{figure*}

We show in \Fig{fig:test_JO424} the results at three different times, $t=0, 400, 600,$ and $800\,\rm s$, comparing \ramses{} in solid curves against \athena{} in dotted curves. The leftmost column of plots shows the fiducial fixed $\ccr=200 \, \cms$ case. The two codes show  very good agreement. The only discernible difference is in the CR flux, which is slightly lower close to the left boundary and beyond the cloud with \ramses{} than \athena{}. This discrepancy mostly disappears if we use a fixed resolution of $2048$ cells in \ramses{} (not shown). These results are well converged with increasing $\ccr$. However, we cannot arbitrarily decrease $\ccr$. In the middle column of plots we use a fixed $\ccr=3 \, \cms$ in \ramses{}. For reference we compare to the same \athena{} run with $\ccr=200 \, \cms$ (but note that using $\ccr=3 \, \cms$ with \athena{} yields very similar results as \ramses{} here). The results are heavily affected by the use of a small $\ccr$, which here is slightly smaller than the highest Alfv\'en velocities. In the rightmost column of plots we run \ramses{} with our new variable $\ccr$ approach. Here we typically get $\ccr\approx50 \, \cms$ (from Eq.~\ref{eq:cvar}), so the CR propagation keeps up with the streaming velocity, and a converged result is retrieved, in very good agreement with the \athena{} result (shown here again using a fixed $\ccr=200 \, \cms$). The variable $\ccr$ case naturally gives a large speedup, completing in 5 seconds on 4 CPUs versus 19 seconds for the fixed $\ccr=200 \, \cms$ case. 

\subsection{CR-driven blast wave} \label{sec:2dblast}

We now replicate the 2D CR-driven blast described in \cite{Pakmor16} and later performed in \JO. We set up a square $1\,\rm cm$ wide domain, with $(x,y)=(0,0)$ defined at the centre, containing a homogeneous gas distribution with a density of $\rho=1 \, {\rm g \, cm^{-3}}$, a thermal pressure density of $\Pgas=1.67 \, \Ecrunits$, zero velocity, and a horizontal magnetic field, i.e.~pointing in the $x$-direction, of magnitude $\sqrt{4 \pi} \,\rm G$. The initial background CR energy density is $\Ecr=0.1 \, \Ecrunits$ except for a circular region at the centre with a radius $r=0.02\,\rm cm$, containing a much higher CR energy density of $\Ecr=100 \, {\rm \Ecrunits}$. This central spike in CR energy density at the centre sets off a blast wave.  We use AMR, going from level 6 to 9, i.e.~a coarse resolution of $64^2$ cells and effective fine resolution of $512^2$ cells. We flag cells for refinement when the relative difference between neighbouring cells is more than $5$ percent in gas density, thermal pressure, or CR energy density. We run using a fixed $\ccr=100 \, {\rm cm \, s^{-1}}$, and perform up to 10 CR subcycling steps to fill in each MHD timestep. We perform two runs, one with pure diffusion, where streaming is deactivated and we set $\kcr=0.033 \, \kcrunits$, and another with pure streaming, setting $\kcr=0.033 \times 10^{-6} \, \kcrunits$ and instead activating the streaming-transport and streaming-heating terms. For comparison, we run the same setup with the one-moment implementation in \ramses{} from \cite{Dubois16} and \cite{Dubois19}, once with pure diffusion and once with pure streaming. 

We show the results at $t=0.1\,\rm s$ in \Fig{fig:test_JO423}, with maps of CR energy density, gas density, thermal pressure, and refinement level, from left to right. We plot contours, as indicated in the color bars, for the CR and gas quantities, with the two-moment results in solid thin curves and the one-moment results in thicker dashed curves. For the non-streaming case, the overlap of the solid and dashed contours is almost perfect, whereas for the streaming case, the agreement is not as good.  We note that the results are almost identical to using a full fixed $512^2$ resolution, with and without streaming, indicating that using AMR works very well.

We have also compared with the same setup in the \athena{} code, and we find excellent agreement with our two-moment implementation in \ramses{}, in both the pure diffusion and pure streaming cases. For completeness, we demonstrate this excellent agreement with \athena{} in \App{appendix:2dblast_athena}.

We have performed the same runs, not shown here, with diffusion and with streaming, using the variable $\ccr$, floored at $\ccrmin=10\,\cms$. Those give almost identical results as shown in \Fig{fig:test_JO423}. Constraining $\ccr \ge 10 c_{\rm MHD}$ (see Eq.~\ref{eq:cvar}), about a third of the number of steps is needed to complete the simulations compared to the fixed $\ccrmin=200\,\cms$ case (and hence the run is three times faster). Additionally constraining $\ccr$ to be larger than 10 times the diffusion speed gives a significantly longer runtime in the pure diffusion case, since the diffusion velocity is fairly high, on the order of $20\,\cms$ at the finest level, but no change in the pure streaming case.

\begin{figure*}
  \centering
  \subfloat
  {\includegraphics[width=0.3\textwidth]
    {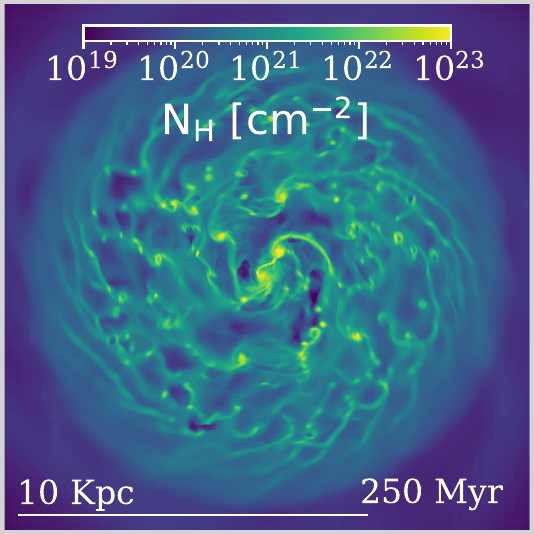}}
  \subfloat
  {\includegraphics[width=0.3\textwidth]
    {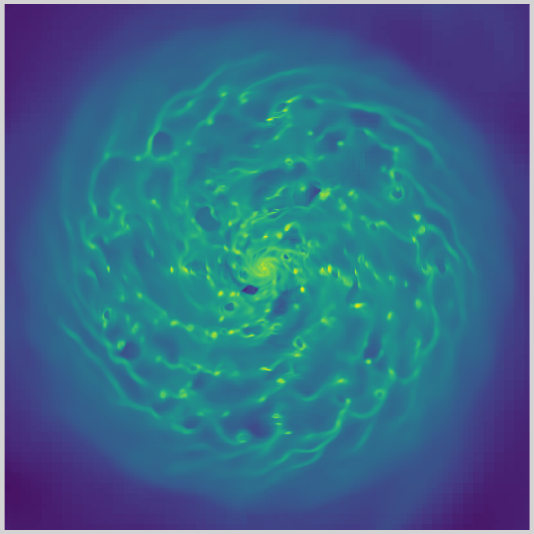}}
  \subfloat
  {\includegraphics[width=0.3\textwidth]
    {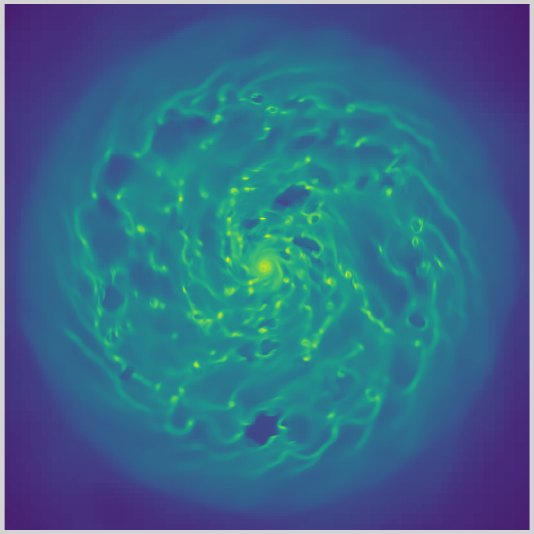}}\\
  \vspace{-4mm}
  \subfloat
  {\includegraphics[width=0.3\textwidth]
    {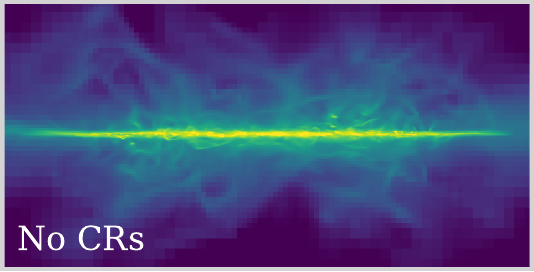}}
  \subfloat
  {\includegraphics[width=0.3\textwidth]
    {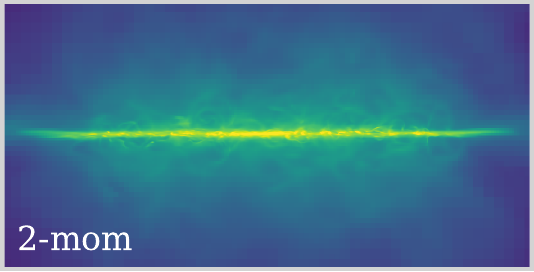}}
  \subfloat
  {\includegraphics[width=0.3\textwidth]
    {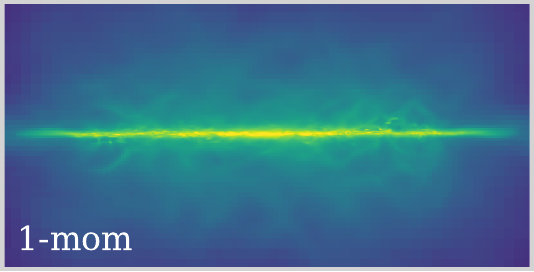}}\\
  \hspace*{5.25cm}
  \subfloat
  {\includegraphics[width=0.3\textwidth]
    {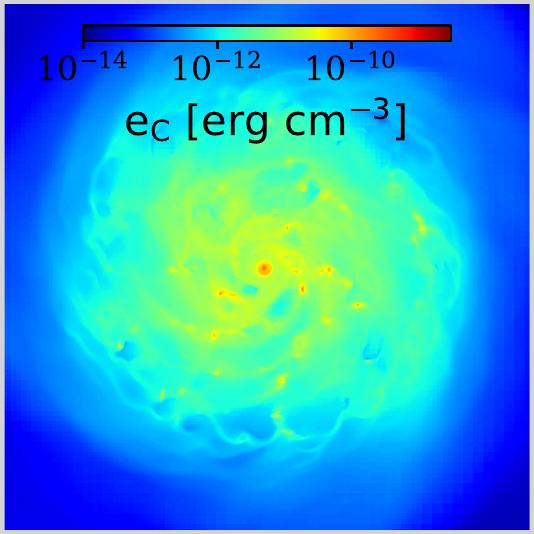}}
  \subfloat
  {\includegraphics[width=0.3\textwidth]
    {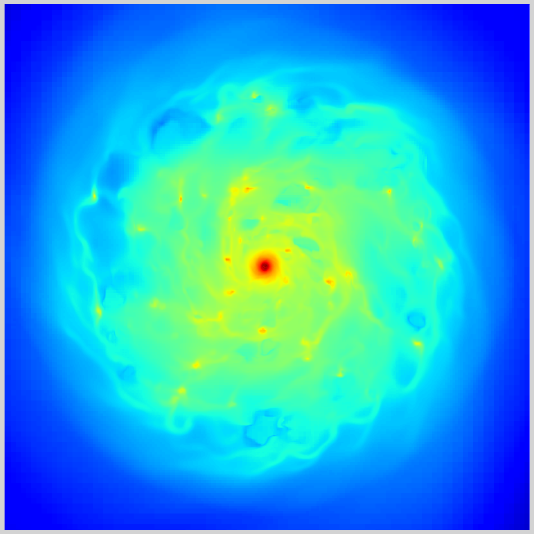}}\\
  \vspace{-4mm}
  \hspace*{5.25cm}
  \subfloat
  {\includegraphics[width=0.3\textwidth]
    {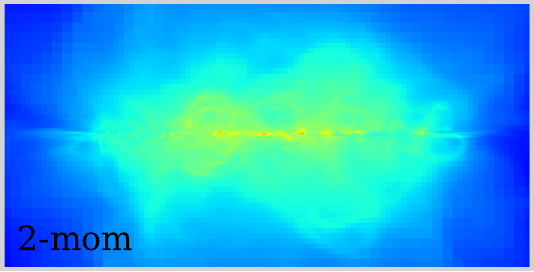}}
  \subfloat
  {\includegraphics[width=0.3\textwidth]
    {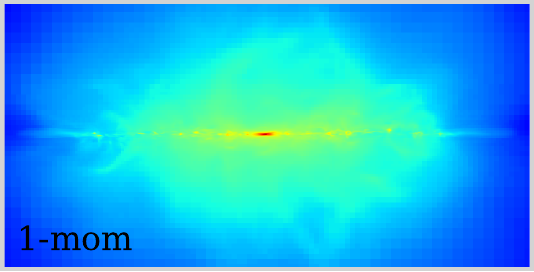}}
  \caption
  {\label{fig:g9_maps} Face-on and edge-on maps from the isolated galaxy runs at at $250\,\rm Myr$. We show hydrogen column densities, $N_{\rm H}$, and CR energy densities, $\Ecr$, for the runs without CRs, with two-moment CRs, and with one-moment CRs, as indicated with labels. The CR energy densities are gas density weighted averages along the projection. The two-moment and one-moment methods produce qualitatively similar gas densities and CR energy distributions, though both the gas and CRs are somewhat more centrally concentrated in the one-moment case.}
\end{figure*}

\section{Isolated galaxy with cosmic rays}\label{sec:galaxy}
We finally test the two-moment CR implementation in \ramses{} on a galactic disc setup where CRs are injected with supernova explosions and play a strong role in shaping the properties of the galactic outflows. We test here whether the new two-moment implementation in \ramses{} gives similar results in suppressing star formation and generating outflows as the already-existing one-moment implementation in the same code, described by \cite{Dubois16}.

\subsection{Initial conditions and subgrid physics}
The initial conditions are the same as the 'G9' galaxy with CR feedback from~\citet{Dashyan20} and~\citet{Farcy22}. We start with a disc of gas and stars plus a bulge of stars embedded in a large-scale dark matter (DM) halo. The disc density follows an exponentially decreasing profile in circular rotation, the bulge (10\% of the total stellar mass) has a Hernquist profile, and the halo has an NFW profile, with the initial conditions produced by {\sc makedisc}~\citep{Springel05} sampled with $N_{\rm d}=10^6$, $N_{\rm b}=10^5$, and $N_{\rm h}=10^6$ particles, for the stellar disc, bulge, and DM halo respectively. Stars from the initial conditions do not contribute to the feedback. The galaxy has a halo mass of $10^{11}\,\rm M_\odot$, and a galaxy baryonic mass of $3.5\times 10^{9}\,\rm M_\odot$. The gas fraction (gas mass over baryonic mass) is 50\% and the metal mass fraction of both gas and stars is initialised at $0.002$, i.e.~about a tenth of Solar metallicity. The simulation starts with a large-scale toroidal magnetic field, with a magnitude proportional to $\rho^{2/3}$ and of the order of $1\,\rm \mu G$ in the galaxy.

The simulation box size is $L_{\rm box}=300\,\rm kpc$.  The AMR grid starts at level $\ell_{\rm min}=7$ and is refined to $\ell_{\rm max}=14$, corresponding respectively to cell sizes of $2.34\,\rm kpc$ and $18\,\rm pc$. Cells are flagged for refinement using a pseudo-Lagragian refinement criterion: a cell is flagged for refinement if the total mass contained in it is larger than $8\times 10^5\,\rm M_\odot$, or if the local Jeans length is smaller than four cell widths.

The simulations include gas cooling down to $10\,\rm K$ using~\cite{Sutherland93} and~\cite{Rosen95} cooling functions. Gas is allowed to form stars following a Schmidt law at a star formation mass density rate of $\dot \rho_\star=\varepsilon_\star \rho/t_{\rm ff}$, where $t_{\rm ff}$ is the free-fall time of the gas and $\varepsilon_\star$ is a gravo-turbulent variable star formation efficiency, which depends on the turbulent Mach number and the virial parameter (see~\citealp{Farcy22} for details). Individual massive stars explode over 3 to $50\,\rm Myr$ into type II core-collapse supernovae (SNII) releasing a fraction $f_{\rm c,SN}=0.1$ of their $10^{51} \,\rm erg$ into CR energy, while the rest of the energy is released in a thermal component following the numerical strategy for momentum conservation from~\cite{Kimm14}.

The simulations are run using anisotropic diffusion with $\kappa=10^{28}\,\kcrunits$, and without streaming. We do not include CR cooling losses, i.e.~$\mathcal{L}_{\rm rad,c}=0$. This is unphysical and leads to a stronger effect of CR feedback than reported in previous works with the same setup, which did include CR cooling, but it is the appropriate choice in our case to simplify as much as possible the comparison between different implementations. As everywhere else in the paper, we assume that CRs are a relativistic population of protons, hence $\gamma_{\rm c}=4/3$.

We run a suite of 3 MHD simulations, each run for $500\,\rm Myr$:
\begin{itemize}
    \item without CRs (i.e.~$f_{\rm c,SN}=0$), referred to as \runnocr;
    \item with CRs and the two-moment implementation of this work, referred to as \runtwomom;
    \item with CRs and the one-moment implementation of~\cite{Dubois16}, referred to as \runonemom.
\end{itemize}

In the \runtwomom~run, we subcycle the CR solver up to 10 times per MHD step (i.e.~$\Nsc=10$) and we use the variable $\ccr$, floored at $\ccrmin=10^{-3}c$ (the actual value of $\ccr$ fluctuates between typical values 10 and 100 times larger than that throughout the simulation). We also gradually decouple CRs from gas when the hydrogen number density becomes lower than $3\times10^{-5} \, {\rm cm^{-3}}$. This is a purely artificial remedy to prevent high velocities and sound speeds in very diffuse gas, which leads to small timestep lengths and hence long computation time. These decoupled gas densities are well below the typical densities in the outflows, so the CR-gas decoupling at low densities does not affect the results. We include this decoupling density as a \ramses{} setup variable, with a default value of $10^4$ times the minimum allowed gas density. A similar feature is used in the \runonemom~run, to alleviate the same issue: the CR energy density is limited locally so that the CR ``temperature'' $T_{\rm c}=(m_{\rm p}/k_{\rm B})(\Pcr /\rho)$ is less than $10^8\,\rm K$.

\begin{figure}
  \centering
  {\includegraphics[width=0.4\textwidth]
    {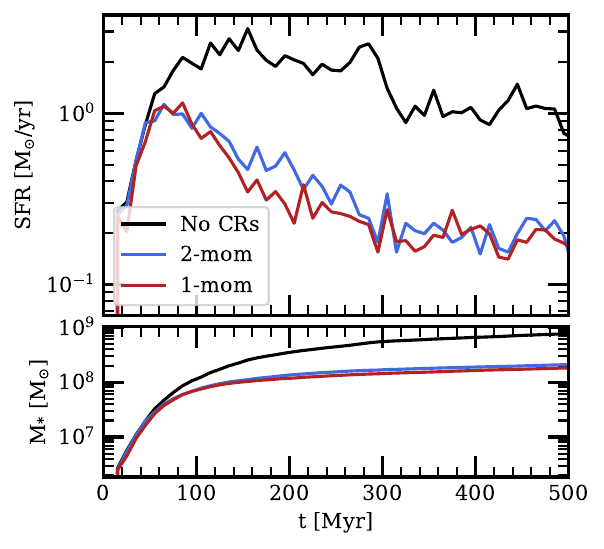}}
  \caption
  {\label{fig:g9_sfr}Star formation rates (upper panel) and stellar mass formed (lower panel) for the isolated galaxy runs without CRs, with two-moment CRs, and with one-moment CRs, as indicated in the legend. The two-moment method gives a similar, though slightly less efficient, suppression of star formation compared to the one-moment method.}
\end{figure}

\begin{figure}
  \centering
  {\includegraphics[width=0.4\textwidth]
    {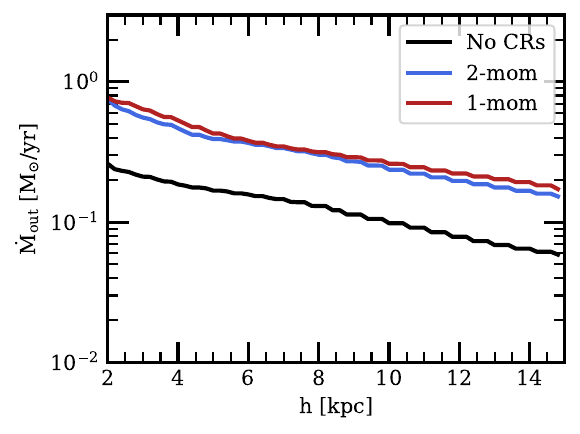}}
  \caption
  {\label{fig:g9_outflow}Gross outflow rates, i.e.~without subtracting inflow rates, averaged for snapshots at 50-Myr intervals between 200 and $500\,\rm Myr$, as a function of height $h$ from the galactic plane, for the isolated galaxy runs without CRs, with two-moment CRs, and with one-moment CRs, as indicated in the legend. The two-moment CRs produce somewhat smaller outflow rates than the one-moment method, but still much higher than without CRs.}
\end{figure}

\subsection{Results for wind launching and star formation rate}
We make a qualitative comparison between the \runnocr, \runtwomom, and \runonemom{} runs in \Fig{fig:g9_maps}, which shows face-on and edge-on maps at $250\,\rm Myr$ of hydrogen column densities, $N_{\rm H}$, and gas-density-weighted averages of the CR energy density, $\Ecr$. The gas densities are similar between the \runtwomom{} and \runonemom{} runs and clearly different from the \runnocr{} case, with less distinct gas clumps, smaller contrasts in density, and smoother distributions, as well as a denser circum-galactic medium.  These effects of CR feedback are generally found in galaxy-scale simulations. The only significant difference between the \runtwomom{} and \runonemom{} runs is that the latter has slightly more concentrated gas at the centre of the galaxy. This difference is not a transient one, but persists more or less throughout the $500\,\rm Myr$ runtime. Similar conclusions can be drawn from the comparison of the $\Ecr$ maps for the two simulations with CRs: the two are fairly similar, but the \runonemom{} run has a significantly stronger concentration of CRs at the galaxy centre (this is also true throughout the run).

We then consider the effect of the two implementations on the formation of stars, and compare to the \runnocr{} case. We show in \Fig{fig:g9_sfr} the star formation rate, SFR, averaged over a timescale of $10\,\rm Myr$, as a function of time, as well as the cumulative stellar mass formed. Both CR implementations have a similarly strong effect in suppressing star formation by a factor of about 5 compared to the SN-only feedback in the \runnocr{} run. The \runonemom{} run has a marginally stronger suppression than the \runtwomom{} one.

We finally look at outflow rates in \Fig{fig:g9_outflow}. We consider outflow rates across planes parallel with the galaxy disc at different heights, and we calculate gross outflow rates, i.e.~taking into account only outflowing gas and not subtracting inflowing gas. The outflow rates at each height are averages from 7 snapshots at $50\,\rm Myr$ intervals from 200 to $500\,\rm Myr$. As for the SFRs, both CR implementations produce a strong boost in the outflow rate compared to the \runnocr{} case, with marginally stronger outflows in the \runonemom{} case. Commenting on the density and velocity profiles of the outflows (not shown): the outflow densities are higher in the CR runs than in the \runnocr{} case by almost an order of magnitude at any distance considered, and the densities are marginally lower (by $\la 5$ percent) in the \runtwomom{} run compared to \runonemom{}. Mass-weighted outflow velocities are lower by a factor 2-3 in the CR runs compared to the \runnocr{} one, and the velocities are marginally higher in the \runtwomom{} case than in the \runonemom{} one.

The capping of local wave speeds using the effective optical depth is vital in the \runtwomom{} run. Due to the highly clustered SN feedback, the simulation produces very high MHD wave speeds in diffuse gas of order 1 percent of the (non-reduced) light speed. This gives effective optical depths, from \Eq{eq:tau}, with values typically between 10 and 50, depending on the varying timestep length. This strongly caps the signal speed, limiting the numerical diffusion of CRs and profoundly affecting the star formation and outflow rates. We have run an equivalent simulation to \runtwomom{} with $\tau=0$ everywhere. Here the artificial diffusion due to zero effective optical depths leads to a much weaker effect of CRs, with a stellar mass of $6\times 10^8 \ M_{\odot}$ formed at $500$ Myr, which is 3 times larger than in the \runtwomom{} run and only slightly smaller than in the \runnocr{} run, and outflow rates at 10 kpc from the disk that are $\approx 50$ percent lower than in the \runtwomom{} run. Even using the \JO{} formulation of $\tau_{\rm JO}$, which gives a factor $2/3$ smaller values than \Eq{eq:tau}, has a non-negligible effect on the results, with 10 percent more stellar mass formed and 20 percent less outflows than in the fiducial \runtwomom{} run.

We note that we have performed an analogue of the \runtwomom{} run with CR streaming, but it a has negligible effect on the results in agreement with the results of~\citet{Dashyan20} for a similar setup using the \runonemom{} implementation.

\section{Conclusions} \label{sec:conclusions}

We have presented a new implementation in the \ramses{} AMR code of the two-moment formulation of CR transport~\citep[e.g.][]{Jiang18,Thomas19,Hopkins22} coupled to MHD. 
CR transport includes diffusion and streaming with any arbitrary level of anisotropic transport along the magnetic field lines together with advection with the gas.
We use an IMEX time-integration procedure to evolve the CR energy density and its flux in every AMR grid cell under this system of hyperbolic equations. The method avoids very small timestep lengths by using the reduced speed of light approximation, i.e.~setting an upper limit on any signal speeds for CRs that is significantly smaller than the light speed. An implicit one-moment CR solver already exists in \ramses~\citep{Dubois16,Dubois19}, but the new two-moment solver had advantages over it in being less numerically diffusive and being trivially able to deal with non-linear diffusion terms (such as streaming) and entropy-violating conditions stemming from anisotropic diffusion without significant additional cost.

Our implementation is closely built on that from \JO{} \citep{Jiang18} in the \athena{} code, except for the following differences:
\begin{itemize}
    \item For increased stability of the method, we use the Lax-Friedrich inter-cell flux function, which is slightly more diffusive than the  Harten–Lax–van Leer one used in \JO, and for momentum injection from CRs to gas, we use the gradient of the CR pressure directly. These modifications eliminates severe instabilities across shock discontinuities that otherwise appear when the optical depth to CRs over the cell width becomes significantly larger than unity.
    \item We use a slightly different formulation for optical depths $\tau$ of cells to CRs, which limit the signal speed when the CRs start to become trapped. Our formulation is based on the ratio of the diffusion time through the cell and the timestep length, and gives roughly a factor $3/2$ higher optical depths compared to \JO.
    \item To reduce the computation time, we subcycle the CR transport and gas-interactions over the MHD timestep, independently at each AMR level, in combination with Dirichlet boundary conditions between refinement levels. This comes with the sacrifice of perfect CR energy conservation between levels.
    \item Instead of a fixed reduced light speed $\ccr$, which forces a very conservative choice of value and hence unnecessarily slow runtimes, we use a variable light speed. Here $\ccr$ adjusts on-the-fly throughout the run so that it is always several times faster than the maximum gas speed (sound speed, MHD wave speed, or velocity), and optionally faster than streaming and diffusion speeds.
\end{itemize}

We have presented extensive classical tests of CR transport, comparing to analytic results where they exist and otherwise to other implementations. This includes one-dimensional and multi-dimensional tests, pure diffusion, pure streaming, or fully CRMHD tests including strong shocks such as in an expanding SNR. This numerical implementation behaves well in all cases. We have highlighted pitfalls in some of the tests where the wrong design choices (not affecting our fiducial method) can lead to bad behaviour and trigger severe numerical instabilities. We have additionally performed isolated galaxy simulations with CR feedback, where CRs are injected in SNRs, showing the typical CR feedback effects of suppressed star formation and increased outflows, in line with expectations from the literature and very similar to the effects obtained with the previously existing one-moment implementation in \ramses.
Overall, we showed that the two-moment method accurately recovers the solution of the traditional one-moment approach as long as the expected solution is close to flux steady-state, and that the two methods lead to nearly equivalent system properties even for complex galaxy feedback situations.

The method can be generalized to include the evolution of gyro-resonant Alfvén wave energy densities, as in~\cite{Thomas21}, or more complex closure relations of the gyroresonant distribution function represented by the $\vec\nabla(\mathcal D_{\rm c} e_{\rm c})$ term in Eq.~\ref{eq:fecr}~\citep{Thomas22,Hopkins22closure}, though we leave these extensions to future work.
A multigroup two-moment scheme is currently under development (Diallo et al., in prep.) to capture the momentum-space dependence of the CR distribution function, a crucial ingredient since CR diffusion coefficients depend on particle energy.
Finally, the two-moment approach offers a natural framework to recast parabolic transport processes, such as thermal conduction or viscosity, into hyperbolic form. In Appendix~\ref{appendix:conduction}, we present the two-moment formulation of thermal conduction together with validation tests in \ramses.

We intend to make the new implementation a part of the public \ramses{} repository\footnote{\url{https://github.com/ramses-organisation/ramses/}} in the near future. In the meantime, we will readily share the code upon request.

\section*{Acknowledgements}

We thank Marion Farcy, San Han, Yan-Fei Jiang, Christoph Pfrommer, Francisco Rodr\'iguez Montero, Matt Sampson, Romain Teyssier, and Timon Thomas for useful discussions.
We gratefully acknowledge support from the CBPsmn (PSMN, Pôle Scientifique de Modélisation Numérique) of the ENS de Lyon for the computing resources. The platform operates the SIDUS solution \citep{Quemener13} developed by Emmanuel Quemener. This work has been supported by the CNRS/INSU Astro French National Programs (co-funded by CEA, CNES, IN2P3, and INP): "Action Thématique de Cosmologie et Galaxies" (ATCG), "Action Thématique de Physique Stellaire" (ATPS), “Action Thématique Phénomènes Extrêmes et Multi-messagers” (AT-PEM), and "Action Thématique Physique et Chimie du Milieu Interstellaire" (AT-PCMI). This work has also made use of the Infinity Cluster hosted by Institut d’Astrophysique de Paris; we thank Stéphane Rouberol for smoothly running this cluster for us. For the purpose of open access, the author has applied a Creative Commons Attribution (CC BY) licence to any Author Accepted Manuscript version arising from this submission.

\cleardoublepage
\appendix

\section{Streaming: various formulations}
\label{appendix:streaming_vel_vs_damping}

\begin{figure}
  \centering
  \subfloat
  {\includegraphics[width=0.37\textwidth]
    {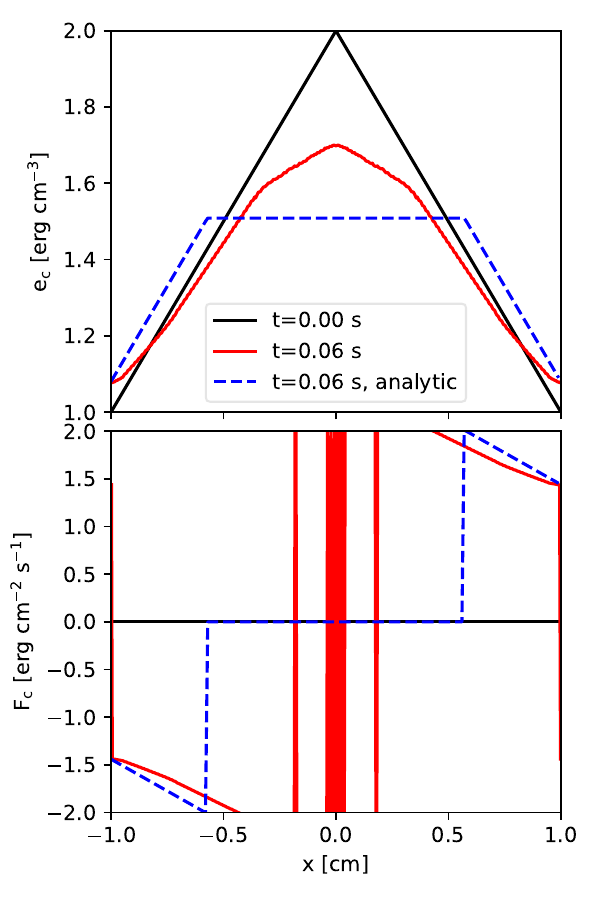}}
  \caption
  {\label{fig:test_JO411_tri_vstream}Streaming test, from \Sec{sec:triangle_test}, with the streaming velocity in the flux source term, i.e.~as in \Eq{eq:fecr}, rather as in \Eq{eq:fecr2}. The solution becomes much closer to the analytic one with $\ccr=10^3\,\cms$ (here it is $\ccr=100\,\cms$), but at the cost of 10 times longer runtime, and even then the agreement is much worse than without the streaming flux source term and $\ccr=100\,\cms$.}
\end{figure}

The transport by the streaming instability appears as $-\sigma_{\kappa}(\vec F-(u+u_{\rm s})(e_{\rm c}+P_{\rm c}))$ in the correct CR flux equation (Eq.~\ref{eq:fecr}). We show the result of the simple streaming test of an initial triangular distribution of CR energy density for this particular formulation of the CR equations in Fig.~\ref{fig:test_JO411_tri_vstream}.
This shows that using this particular implementation of streaming transport the solution on $e_{\rm c}$ is wrong unless using extremely large values of the reduced speed of light.A

\section{Static 1D test}
\label{appendix:static}
\begin{figure}
  \centering
  \subfloat
  {\includegraphics[width=0.40\textwidth]
    {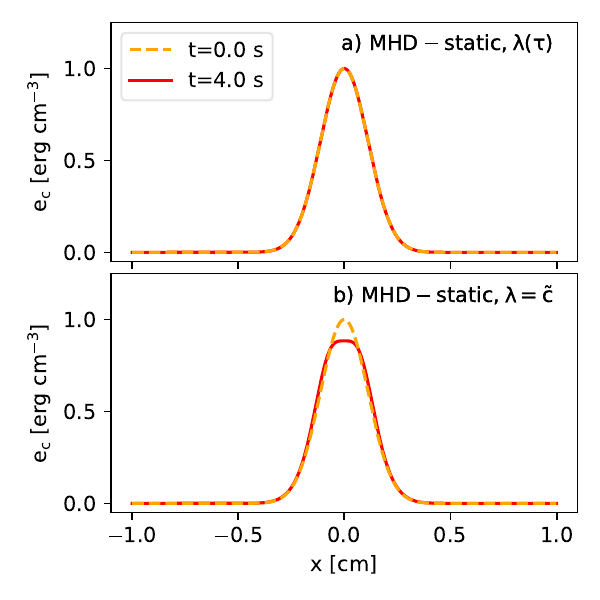}}
  \caption
  {\label{fig:test_JO414_tauzero}Tests of the non-diffusion of a Gaussian CR energy distribution, which should not change from the initial conditions. Panel {\bf a)} shows the test run with our fiducial CR solver, limiting the signal speed $\lambda$ locally by the diffusion time and gas velocity, which here are both essentially zero. In this case there is no diffusion of CRs. Panel {\bf b)} shows the run with the only modification that $\lambda =\tilde c$ (with $\tilde c=100\,\cms$).}
\end{figure}

In this test, we wish to show the importance of limiting the CR signal speed with the effective optical depth $\tau$ (Eq. \ref{eq:tau}). We start an MHD-static run from a Gaussian distribution of the CR energy density, as desribed in \Sec{sec:diffusion_test} and deactivate all the physical transport terms of the CR quantities, with $\kappa=10^{-15} \, \rm cm^2 \, s^{-1}$, no streaming, and zero gas velocity. The Gaussian is, hence, supposed to stay identical to its initial conditions, and, the only difference might result from numerical diffusion in the numerical solver of the CR two-moment equations. In \Fig{fig:test_JO414_tauzero}, we compare the results at $t=4\,\rm s$ for our fiducial method, in panel a), with a version of the conservative update of the two-moment equations using $\lambda =\tilde c$ instead of $\lambda$ being limited by the diffusion speed or gas velocity ($=0$ in this test). The numerical diffusion in panel b) demonstrates the importance of limiting the signal speed.

\section{Non-static 1D test} \label{sec:stationary_test}
The optically thick regime is prone to numerical instabilities when the MHD response of the gas to cosmic rays is activated. We repeat here the 1D diffusion test from above, with a near-zero diffusion coefficient, $\kcr =  10^{-15} \, \kcrunits$, zero gas velocity, and here a fixed resolution level $\ell=9$, corresponding to 512 cells spanning the domain.

\begin{figure}
  \centering
  {\includegraphics[width=0.37\textwidth]
    {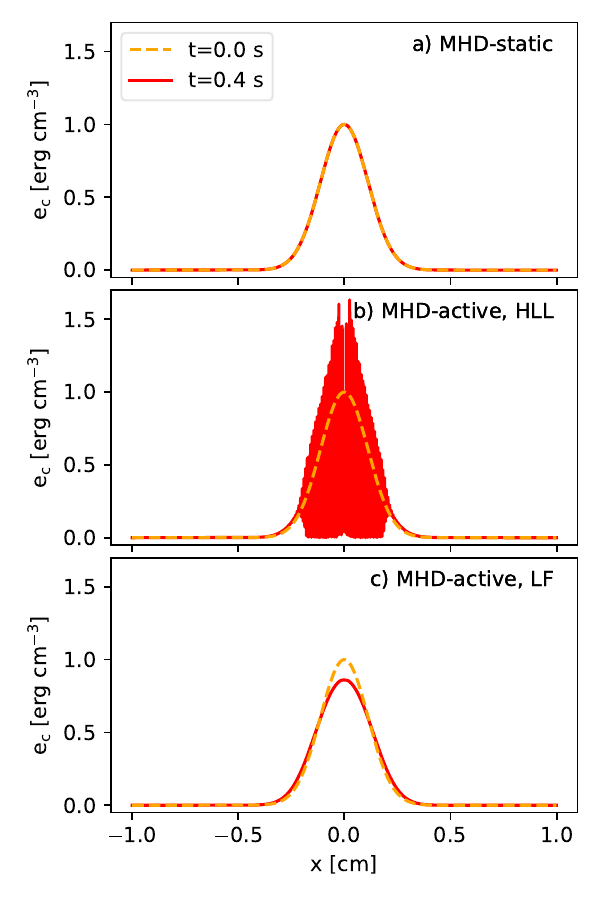}}
  \caption
  {\label{fig:test_JO414_nodiff}From top to bottom: {\bf a)} Stationary CR test (i.e. with MHD turned off). The CR distribution stays in place, as it should. {\bf b)} CR test with MHD turned on, and here using a momentum injection from CRs that mirrors the source term of the CR flux equation combined with the HLLE inter-cell flux function (see text for details). This approach leads to numerical instabilities in the CR distribution. {\bf c)} CR test with MHD turned on, using our fiducial implementation with the LF inter-cell flux and directly using the CR pressure gradient for momentum injection. The  CR pressure pushes the gas, and the work done by the  CRs leads to a loss of CR energy. More importantly, the code is stable in this case.}
\end{figure}

We first show the results with static MHD in panel a) of \Fig{fig:test_JO414_nodiff}. In this case the code behaves well thanks to the signal speed $\lambda$ of the inter-cell flux function being modulated by the effective optical depth $\tau$.

The stability of the numerical method becomes more complex when activating the MHD, allowing the gas to react to the CR pressure. We first illustrate in panel b) of \Fig{fig:test_JO414_nodiff} the behaviour if we use the non-fiducial combination of $\sigma(\vec F_{\rm c}-\vec u(e_{\rm c}+P_{\rm c}))$ for CRs to gas momentum injection (instead of $-\vec \nabla P_{\rm c}$) and an HLLE inter-cell flux. The run develops strong numerical instabilities. Our fiducial approach to avoiding those instabilities, which tend to appear when $\tau\gg1$, is to use the local CR pressure gradient for CR injection, as shown in Eq.~\ref{eq:momentum2}, and the LF inter-cell flux function when for the CR conservative step. The test results when applying these modifications are shown in panel c) of \Fig{fig:test_JO414_nodiff}. Here the results are free of these unwanted numerical instabilities and we simply see a small reduction in CR energy due to the work done pushing the gas.

We note that either of our modifications i) (LF inter-cell flux) or ii) (CR pressure gradient momentum injection) is sufficient to avoid instabilities in this test. We have decided to use both modifications by default in our implementation, as this generally prevents small oscillatory features in this test and others, which  can appear when only one of the modifications is applied. 

\section{CR-driven blast wave: comparison with Athena++}
\label{appendix:2dblast_athena}
\begin{figure*}
  \centering
  \subfloat
  {\includegraphics[width=0.99\textwidth]
    {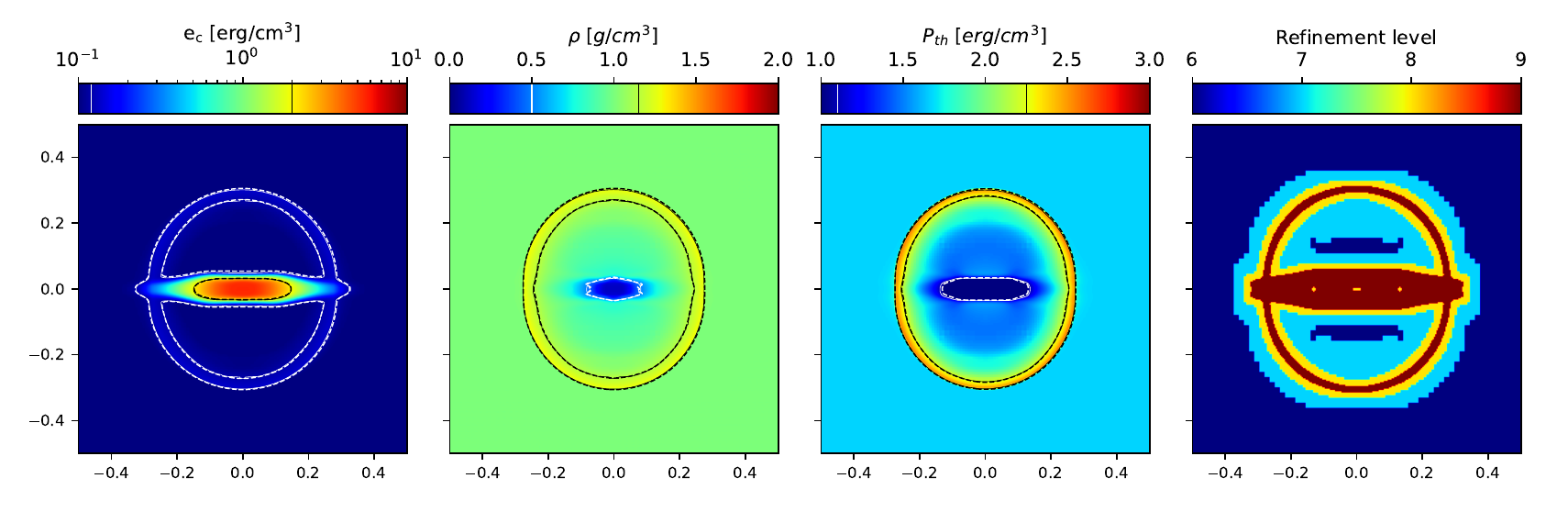}}
  \vspace{-5mm}
  \hspace{-3mm}
  \subfloat
  {\includegraphics[width=0.99\textwidth]
    {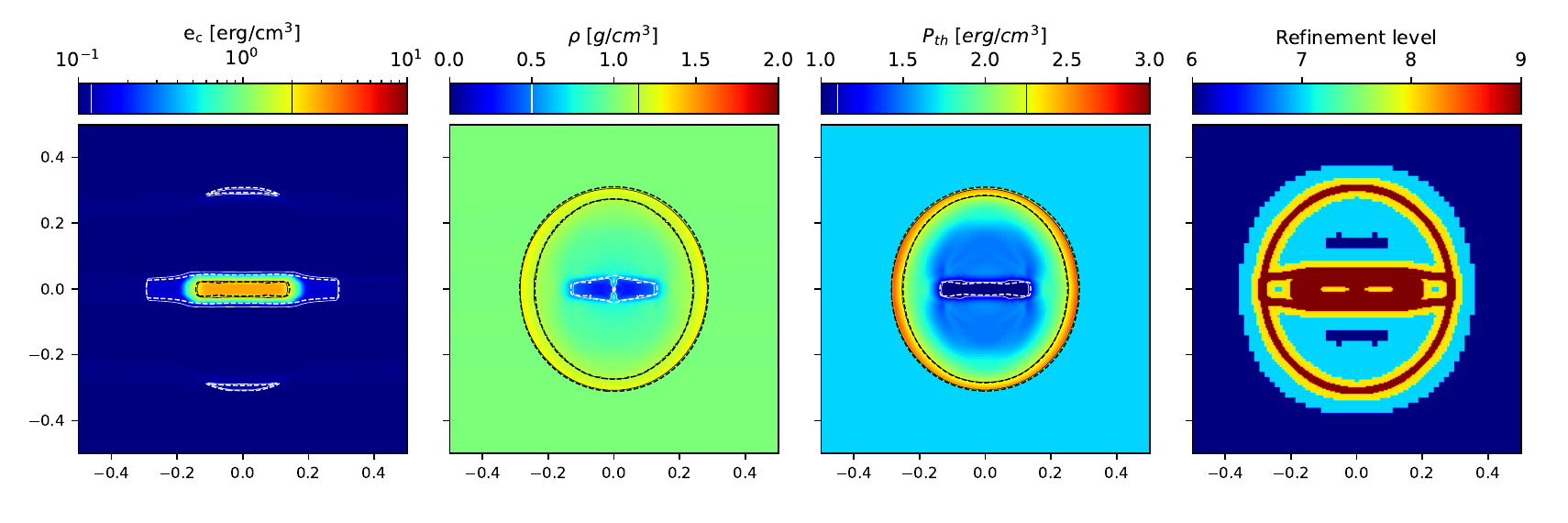}}
  \caption
  {\label{fig:test_JO423_athena} Comparison with \athena{} for the CR-driven blast test. All panels show results at $t=0.1$ s. The upper row is with pure diffusion and the lower row is with pure streaming. From left to right, the panels show CR energy density, gas density, gas thermal pressure, and refinement level (in \ramses; for \athena{} the resolution is fixed at $512^2$ cells). For the gas and CR quantities, we over-plot contours at arbitrarily selected values, as indicated in the color-bars, with solid contours representing the current two-moment method, and dashed contours representing equivalent runs with \athena. The two codes agree very, both for the diffusion and streaming cases.}
\end{figure*}

In \Sec{sec:2dblast} we compared results from a CR-generated two-dimensional blast and compared the results of our two-moment implementation in \ramses{} with that of the one-moment implementation in the same code. Here  we show the same comparison with the \athena{} code. The \athena{} run has the same setup parameters as described in \Sec{sec:2dblast}, except it has a fixed $\ccr=100\,\cms$ and a fixed resolution of $512^2$ cells. We have run it using the 'vl2' integrator. We show the  comparison in  \Fig{fig:test_JO423_athena}, with the map and solid contours representing the two-moment \ramses{} run, just like in \Fig{fig:test_JO423}, and the dashed contours now representing \athena{}. The solid and dashed contours are nearly indistinguishable, demonstrating that the two codes are in excellent agreement.  

\section{CR-driven blast wave with isotropic diffusion}
\label{appendix:2dblast_isotropic}
\begin{figure*}
  \centering
  \subfloat
  {\includegraphics[width=0.99\textwidth]
    {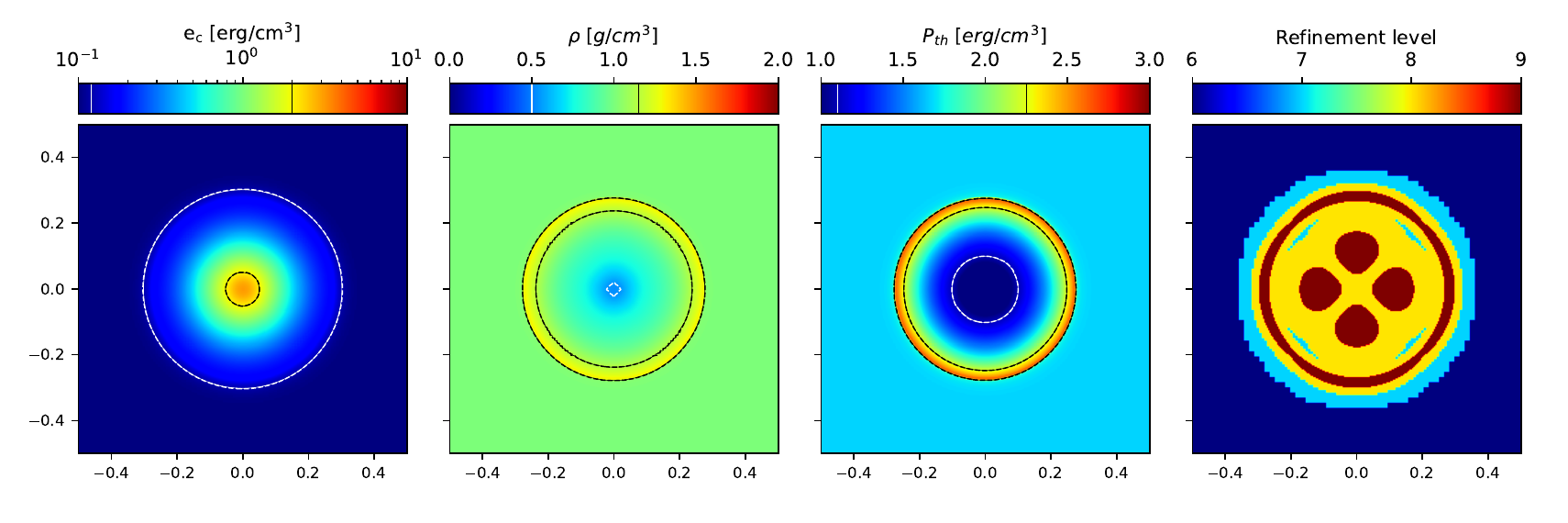}}
  \caption
  {\label{fig:test_JO423_isotropic}CR-driven blast test with isotropic (pure) diffusion and an insignificant magnetic field, but otherwise as described in \Sec{sec:2dblast}. The solid contours trace values as indicated in the color-bars, and for comparison we show in dashed contours results from an equivalent run with the 1-moment method from \protect\cite{Dubois19}. The two are in very good agreement.}
\end{figure*}

In order to demonstrate that isotropic CR propagation works with the current two-moment method, we have run the 2D blast test from \Sec{sec:2dblast} with pure isotropic diffusion, i.e. setting the perpendicular component of $\kappa$ equal to the parallel one (i.e.~$f_{\parallel,\kappa}=0$). Otherwise the setup is the same as described in \Sec{sec:2dblast}, except that we initialise to a thousand times weaker magnetic field magnitude, to minimise magnetic pressure and its impact on shell dynamics (effectively squeezing the aspect of the SNR in the $B$-field perpendicular direction). We show the results in \Fig{fig:test_JO423_isotropic}, where we indeed find an isotropic blast. For comparison we have plotted contours in the three leftmost panels, with the two-moment method represented by solid curves and an equivalent experiment with the one-moment method from \cite{Dubois19} represented by dashed curves. The two are in all cases in excellent agreement and clearly isotropic propagation is no issue with the new method.

\section{Two-moment thermal conduction}
\label{appendix:conduction}

\subsection{Thermal conduction with two temperatures}

Anisotropic thermal conduction can be solved with the same set of non-Fickian equations~\citep{Rempel17,Warnecke20,Navarro22}, following\footnote{This set of hyperbolic heat conduction equations are also called relativistic heat conduction equations or ``Maxwell-Cattaneo-Vernotte'' equations.}:
\begin{align}
& \label{eq:temp} \frac{\partial e_{\rm e}}{\partial t} + \vec{\nabla}  . \vec{F}_{\rm e} = \vec{u}  . \vec{\nabla} P_{\rm e} +\mathcal{C}_{\rm ei}  \,, \\
& \label{eq:felec} \frac{1}{\tilde c^2} \frac{\partial \vec{F}_{\rm e}}{\partial t} + \frac{1}{3} \vec{\nabla} e_{\rm e} = -\sigma_{\rm e} \left( \vec{F}_{\rm e} - \vec{u}(e_{\rm e} + P_{\rm e}) \right) \,,
\end{align}
where $e_{\rm e}$, $P_{\rm e}=(\gamma-1)e_{\rm e}$, $T_{\rm e}$ and $F_{\rm e}$ are respectively the energy density, pressure, temperature and lab-frame heat flux of electrons, and $\mathcal{C}_{\rm ei}$ is the coupling rate of the electron and ion temperatures by small angle Coulomb scatterings.
In that two temperature description of the gas energy, pressure and temperature quoted in the original MHD equations become ion quantities. 
The difference between ion and electron temperatures might become important in a sufficiently hot and diffuse plasma where electron self-collisions are sufficiently important for conduction but electron-ion collisions are too scare to fulfill local thermodynamical equilibrium leading to temperature decoupling.

In the limit $c^{-2}\partial_t F_{\rm e}\rightarrow 0$, this set of equations reduces to:
\begin{equation}
     \frac{\partial e_{\rm e}}{\partial t} + \vec{\nabla}  . (\vec{u} e_{\rm e}) = -P_{\rm e} \vec \nabla.\vec{u} - \vec \nabla \left(-\kappa_{\rm Sp} \vec \nabla T_{\rm e} \right)+\mathcal{C}_{\rm ei}\, ,
\end{equation}
where the conductivity is 
\begin{equation}
     \kappa_{\rm Sp}=n_{\rm e}k_{\rm B}D_{\rm cond}\, , 
\end{equation}
and, hence, $(3\sigma_{\rm e})^{-1}=(\gamma-1)D_{\rm cond}$, with the~\cite{Spitzer56} thermal diffusivity
\begin{equation}
     D_{\rm cond}=8.3\times 10^{28} f_{\rm sat}\left(\frac{T_{\rm e}}{10^8{\rm K}}\right)^{5/2} \left(\frac{n_{\rm e}}{1{\rm cm^{-3}}}\right)^{-1} \,\rm cm^2\, s^{-1}\, ,
\end{equation}
for electrons (assuming a Coulomb logarithm of $\ln \Lambda=40$) with a number density $n_{\rm e}=\rho/(\mu_{\rm e}m_{\rm p})$ ($\mu_{\rm e}\simeq1.14$ for a fully ionised primordial gas).
The thermal diffusion is saturated through the $f_{\rm sat}$ parameter, when the electron mean free path is larger than the length-scale of the temperature gradient~\citep{Cowie77,Sarazin86}, or by the whistler instability~\citep{Roberg16}.
The term $\mathcal{C}_{\rm ei}$ is the coupling term between the ion and electron temperature converging to equilibration by collisions at a rate:
\begin{equation}
\mathcal{C}_{\rm ei}=\frac{(T_{\rm i}-T_{\rm e})}{\tau_{\rm eq}}\frac{n_{\rm e}k_{\rm B}}{\gamma-1}\, ,
\end{equation}
with
\begin{equation}
\tau_{\rm eq}=0.2 \left(\frac{T_{\rm e}}{10^8{\rm K}}\right)^{3/2} \left(\frac{n_{\rm i}}{1{\rm cm^{-3}}}\right)^{-1} \,\rm Myr
\end{equation}
for which we have assumed that ions are only composed of fully ionised hydrogen and helium, with ion number density $n_{\rm i}=\rho/(\mu_{\rm i}m_{\rm p})$ ($\mu_{\rm i}\simeq1.22$ for a fully ionised primordial gas), and used the fact that $m_{\rm p}\gg m_{\rm e}$.

Equations can be rewritten into an hyperbolic transport step $\partial_tq+\nabla f_q=0$, where 
\begin{equation}
q=\begin{pmatrix}e_{\rm e}\\F_{{\rm e},x}\\F_{{\rm e},y}\\F_{{\rm e},z}\end{pmatrix}\ {\rm and} \
f_q=\begin{pmatrix}
 F_{{\rm e},x} & F_{{\rm e},y} & F_{{\rm e},z} \\
 \frac{\tilde c^2}{3} e_{\rm e} & 0 & 0 \\
 0 & \frac{\tilde c^2}{3} e_{\rm e} & 0 \\
 0 & 0 & \frac{\tilde c^2}{3} e_{\rm e}
\end{pmatrix}    \, ,
\end{equation}
and a source step for transport:
\begin{align}
&    \frac{e_{\rm e}^{n+1}-e_{\rm e}^{n}}{\Delta t} = -\vec u.3(\gamma-1)\sigma_{\rm e}^n(\vec F_{\rm e}^{n+1}-\vec u\gamma e_{\rm e}^{n+1}) \, ,  \\
&    \frac{1}{\tilde c^2}\frac{\vec F_{\rm e}^{n+1}-\vec F_{\rm e}^{n}}{\Delta t} = -\sigma_{\rm e}^n\left(\vec F_{\rm e}^{n+1}-\vec u \gamma e_{\rm e}^{n+1}\right)\,, 
\end{align}
where we have exploited the fact that $\tilde c^{-2}\partial_t F_{\rm e}\rightarrow 0$ leads to $\vec \nabla e_{\rm e}/3=-\sigma_{\rm e}(\vec F_{\rm e}-\vec u\gamma e_{\rm e})$.
The coupling term $\mathcal{C}_{\rm ei}$ is dealt with separately using an implicit step~\citep{Dubois16}.

\subsection{Thermal conduction with one temperature}

Using a single gas temperature assumes that equilibration of the temperature/energy of ions and electrons is instantaneous (sufficiently short compared to shock-heating or thermal conduction time scales).
Therefore, one can just handle the divergence of the heat flux\footnote{we call it $F_{\rm heat}$ now for disambiguity but this is akin to $F_{\rm e}$ except it is now in the fluid frame} $\vec \nabla.\vec F_{\rm heat}$ directly in the total energy equation of the plasma, which term will be associated with a flux equation in the \emph{comoving fluid} frame:
\begin{align}
& \frac{\partial e}{\partial t} + \vec{\nabla}  . \left( (e + P) \vec{u} - \frac{(\vec{B}  . \vec{u}) \vec{B}}{4\pi} \right) + \vec \nabla . \vec F_{\rm heat}=0 \,, \\
&\frac{1}{\tilde c^2} \frac{\partial \vec{F}_{\rm heat}}{\partial t} + \frac{1}{3} \vec{\nabla} e = -\sigma_{\rm e}  \vec{F}_{\rm heat} \,.
\end{align}
Therefore, we can partially solve the 
\begin{align}
& \frac{\partial e}{\partial t} + \vec \nabla . \vec F_{\rm heat}=0 \,, \\
&\frac{1}{\tilde c^2} \frac{\partial \vec{F}_{\rm heat}}{\partial t} + \frac{1}{3} \vec{\nabla} e = -\sigma_{\rm e}  \vec{F}_{\rm heat} \, ,
\end{align}
subsystem with our two-moment machinery designed for CR transport, while leaving the first divergence term of the energy equation to the \ramses{} MHD solver.

\begin{figure}
  \centering
  \subfloat
  {\includegraphics[width=0.49\textwidth]
    {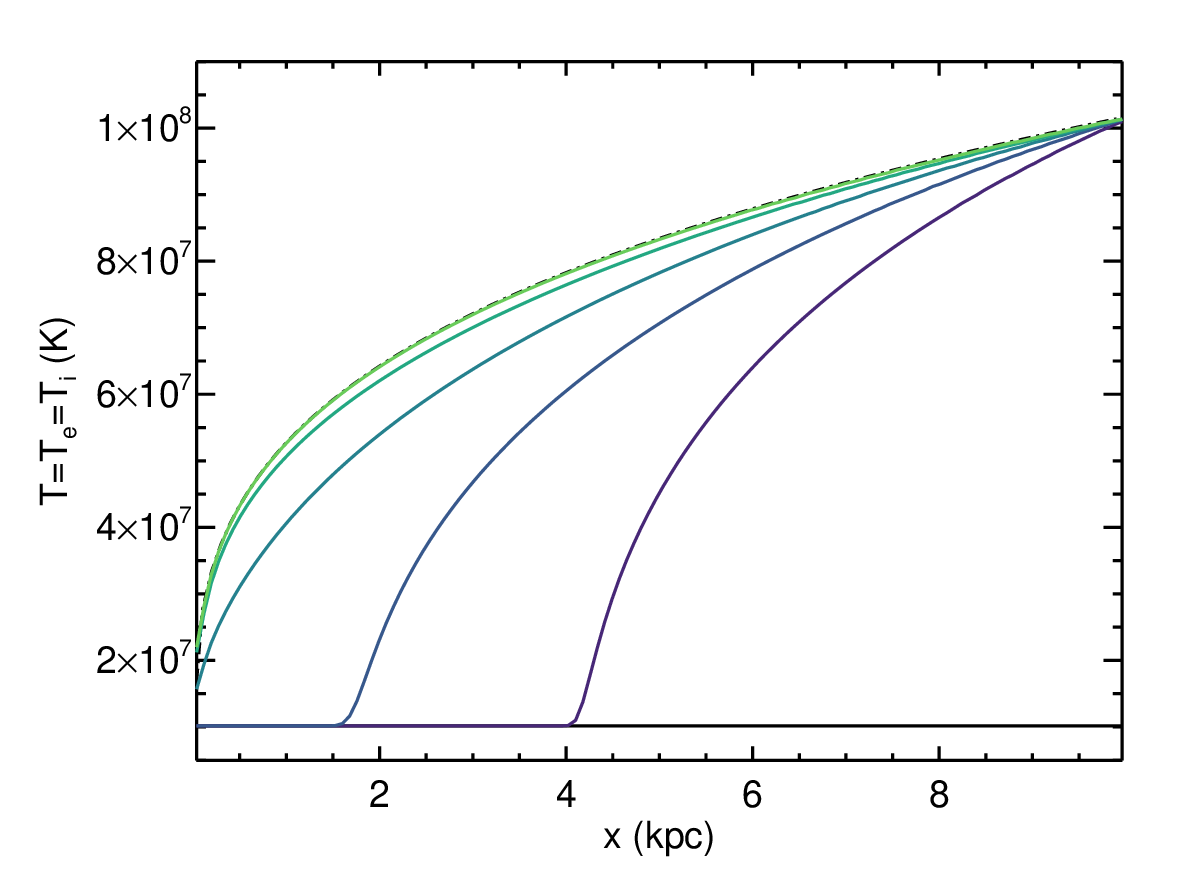}}\vspace{-0.5cm}\\
  {\includegraphics[width=0.49\textwidth]
    {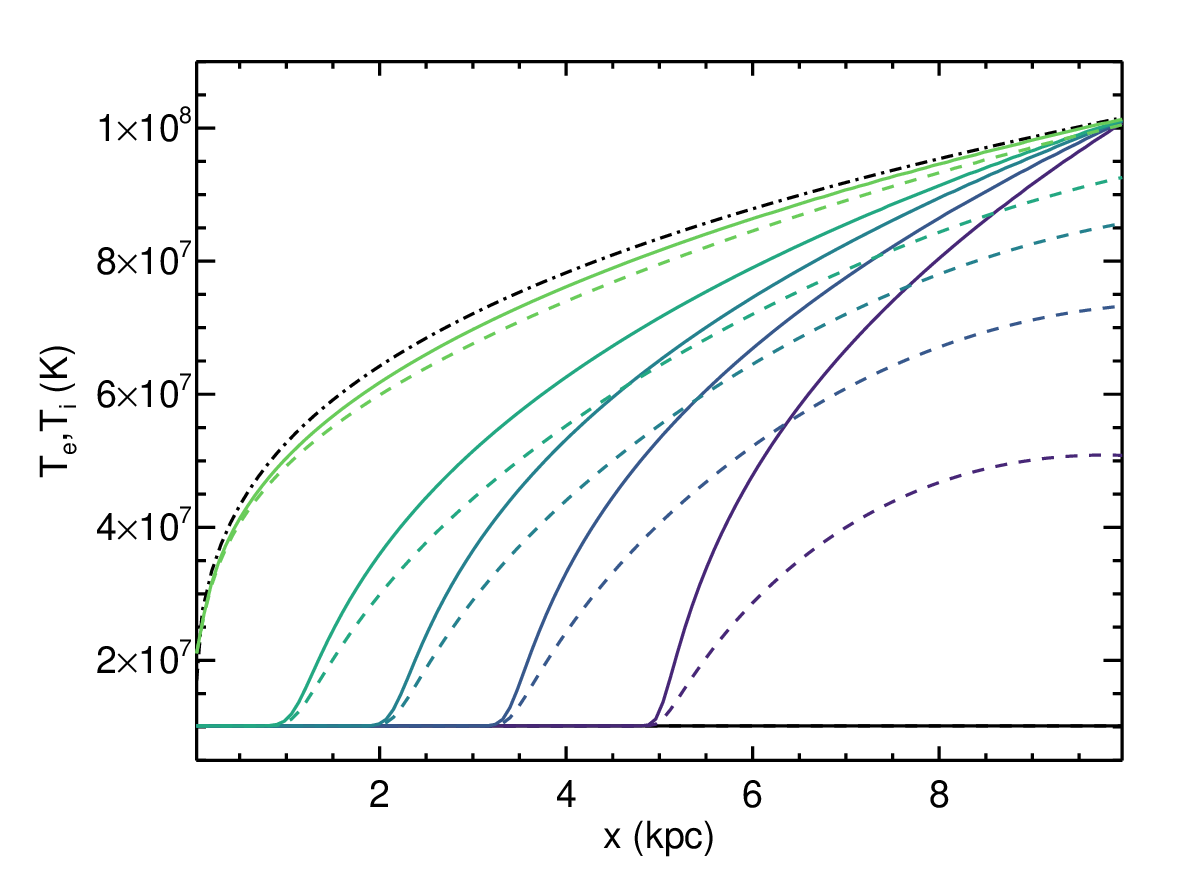}}
  \caption
  {\label{fig:thermal_conduction} Thermal conduction test with Spitzer diffusivity starting from a uniform temperature $T_{\rm e}=10^7\,\rm K$ everywhere except for the right box boundary with $T_{\rm e,R}=10^8\,\rm K$ (black solid line), and its numerical solution obtained with the two-moment method at time $t=2.5,5,7.5,10$, and $20\,\rm Myr$ (from blue to green colors). The top panel corresponds to a single temperature system (i.e.~$T_{\rm e}=T_{\rm i}=T$ by definition), while the bottom panel shows the result for the system with two temperatures ($T_{\rm e}$ in solid and $T_{\rm i}$ in dashed). The steady-state solution in dot-dashed line is well reproduced by the numerical solution, with the two-temperature approach introducing an extra time delay in reaching the steady-state solution.}
\end{figure}

\subsection{Thermal conduction test}

We set a 1D test of the two-moment thermal conduction equations using Spitzer thermal diffusivity and neglecting saturation effects (i.e.~$f_{\rm sat}=1$). We use a box size of $L=10\,\rm kpc$ with 128 cells, a volume filled with $n=10^{-1}\,\rm H\,cm^{-3}$ and $T_{\rm e}(x)= 10^7\,\rm K$, and imposed left and right box boundaries of $T_{\rm e, L}=10^7\,\rm K$ and $T_{\rm e, R}= 10^8\,\rm K$, respectively.
We test the single-temperature system, i.e.~with $T_{\rm e}=T_{\rm i}=T$ by definition, and the two-temperature system for which thermal conduction only applies to electrons and where ions are coupled to electrons through microscopic Coulomb collisions.
The typical time scales for diffusion and temperature coupling for this particular choice of parameters are $\simeq20\,\rm Myr$ and $\simeq2\,\rm Myr$ respectively. 
The steady-state solution $\partial_t T_{\rm e}=0$ for thermal conduction with $D_{\rm c}\propto T_{\rm e}^{2.5}$ is $T_{\rm e}(x)=(T_{\rm e, L}^{7/2}+(T_{\rm e,R}^{7/2}-T_{\rm e,L}^{7/2})x/L)^{2/7}$.

The top panel of Fig.~\ref{fig:thermal_conduction} shows the result of the numerical solution at different times which correctly captures the steady-state solution at long time scales $t\simeq 10\,\rm Myr$.
The bottom panel of Fig.~\ref{fig:thermal_conduction} shows the result of the numerical solution for both the electronic temperature $T_{\rm e}$ and the ionic temperature $T_{\rm i}$ (starting from $T_{\rm i}(x)=T_{\rm e}(x)$).
Decoupling the two temperatures introduces a time delay in reaching the steady-state solution, with ionic temperature slowly catching up with the temperature of the electronic precursor.

\end{document}